\documentclass[aps,preprint,nofootinbib,floatfix]{revtex4-1}
\usepackage{graphicx} 
\usepackage[latin1]{inputenc}
\usepackage{amsmath,amssymb}
\usepackage{hyperref}
\usepackage{epstopdf}
\usepackage{geometry}                
\def\be{\begin{equation}}
\def\ee{\end{equation}}
\def\bea{\begin{eqnarray}}
\def\eea{\end{eqnarray}}
\newcommand{\gev}{~{\rm GeV}}

\newcommand{\mev}{~{\rm MeV}}
\newcommand{\kev}{~{\rm keV}}

\begin{document}

\title{Lepton Flavor Violating Radiative Decays\\ 
in EW-Scale $\nu_R$ Model: An Update}

\author{P. Q. Hung$^{1,4}$}
\email{pqh@virginia.edu}
\author{Trinh Le$^1$}
\email{ttl9ve@virginia.edu}
\author{Van Que Tran$^2$}
\email{apc.tranque@gmail.com}
\author{Tzu-Chiang Yuan$^{3,5}$}
\email{tcyuan@phys.sinica.edu.tw}

\affiliation{
$^1$Department of Physics, University of Virginia, Charlottesville, VA 22904-4714, USA\\
$^2$Department of Physics, National Taiwan Normal University, Taipei 116, Taiwan\\
$^3$Institute of Physics, Academia Sinica, Nangang, Taipei 11529, Taiwan \\
$^4$Center for Theoretical and Computational Physics, Hue University College of Education, Hue, Vietnam\\
$^5$Physics Division, National Center for Theoretical Sciences, Hsinchu, Taiwan
}

\date{\today}                                           

\begin{abstract}

We perform an updated analysis for the one-loop induced lepton flavor violating radiative decays
$l_i \to l_j  \gamma$ in an extended mirror model. Mixing effects of the neutrinos and charged leptons 
constructed with a horizontal $A_4$ symmetry are also taken into account.
Current experimental limit and projected
sensitivity on the branching ratio of $\mu \to e \gamma$ are used to constrain the parameter space of the model.
Calculations of two related observables, the electric and magnetic dipole moments of the leptons, are included. Implications concerning the possible detection of mirror leptons at the LHC and the ILC are also discussed.

\end{abstract}


\maketitle


\section{Introduction}

The electroweak-scale right-handed neutrino (EW-scale $\nu_R$) model was proposed by one of us (PQH) \cite{Hung:2006ap} with the following main motivations in mind: 1) To provide a model for the see-saw mechanism which can be realized at the electroweak scale instead of a typical grand unification theory (GUT) scale; 2) To be able to test the seesaw mechanism through the discovery of right-handed neutrinos whose Majorana masses are now bounded by the electroweak scale $\Lambda_{EW} \sim 246 \gev$; 3) To be able to probe at high energies (e.g. at the Large Hadron Collider (LHC)) lepton-number violating processes such as like-sign dilepton events coming from the Majorana nature of the right-handed neutrinos. The electroweak-scale right-handed neutrinos belong to doublets of the Standard Model (SM) $SU(2)$ whose partners are right-handed ``heavy" mirror charged leptons. The requirement of the absence of anomaly dictates the addition of right-handed doublets of mirror quarks to the particle spectrum. Furthermore, left-handed $SU(2)$-singlet mirror quarks and mirror charged leptons will be the counterparts of their SM right-handed $SU(2)$-singlet quarks and charged leptons.

The EW-scale $\nu_R$ model entails extra $SU(2)$ chiral doublets (the mirror fermions) which have many consequences. These mirror fermions enter loop corrections to various quantities and processes such as the electroweak precision parameters, rare processes, etc.

The first type of effects that needs to be examined is the contributions of these extra chiral doublets to the electroweak precision parameters. These calculations have been performed in \cite{Hoang:2013jfa} and it was found that there is a large parameter space where the EW-scale $\nu_R$ model satisfies the EW precision constraints. In a nutshell, the contributions from the mirror fermions are partially cancelled by those of the scalar sector, in particular the $SU(2)$ triplet scalar.

The next place where mirror fermions enter through loop corrections is rare processes such as $\mu \rightarrow e\, \gamma$ and $\tau \rightarrow \mu \, \gamma$. In \cite{Hung:2007ez}, such processes have been discussed in a generic fashion, with an emphasis on the possible correlation between the observability of the aforementioned rare processes and the decay lengths of the mirror charged leptons, both of which are of phenomenological interests. In this article, we will present an update of the process $\mu \rightarrow e\, \gamma$ taking into account recent developments of the model, including experimental inputs from the recently-discovered 125 GeV SM-like scalar \cite{Aad:2012tfa,Chatrchyan:2012xdj}. They are summarized below.

The scalar sector of the original model \cite{Hung:2006ap}  contains one SM-like Higgs doublet and two Higgs triplets, one with $Y/2=1$ containing doubly-charged scalars and one with $Y/2=0$. (The rationale for this sector will be explained in the summary section.)  The discovery of the 125 GeV SM-like scalar has opened up a whole new chapter on any model beyond the SM, in particular those models which have more than one Higgs doublet.  In light of this discovery, a close examination of the scalar sector of the EW-scale $\nu_R$ model \cite{Hoang:2014pda} revealed that its original Higgs content is insufficient to accommodate the 125 GeV SM-like scalar. It turns out that a simple introduction of an extra Higgs doublet, now totaling two: one of which couples to the SM fermions and the other one to the mirror quarks and charged leptons. This yields two 125-GeV candidates with one being SM-like (dubbed {\em Dr. Jekyll}) and the other being very different ({\em Mr. Hyde}), both of which giving comparable signal strengths in agreement with ATLAS and CMS data.

Most importantly for the present manuscript is the recent work \cite{Hung:2015nva} concerning neutrino and SM charged lepton masses and mixings. The fact that the SM lepton mixing matrix $U_{\rm PMNS}$ 
(the Pontecorvo-Maki-Nakagawa-Sakata mixing matrix)
is so different from the quark counterpart, $V_{\rm CKM}$
(the Cabibbo-Kobayashi-Maskawa mixing matrix), 
has given rise to many models, many of which invoke the presence of some kind of discrete symmetry. Among these different proposals for the discrete symmetry is the popular $A_4$ symmetry which has been used to reproduce the tribimaximal form of $U_{\rm PMNS}$. This symmetry is usually applied to the charged lepton sector \cite{ma} and involves four or more Higgs doublets. (Such a large number of Higgs doublets might be hard to accommodate the 125 GeV SM-like scalar with the desired observed  properties.) The new twist of \cite{Hung:2015nva} is to exhibit the $A_4$ symmetry in the neutrino Dirac mass sector and the scalar sector involved is composed of $SU(2) \times U(1)_Y$-singlet scalars which are not constrained by LHC data. These singlet scalars are composed of a singlet and a triplet of $A_4$. This model reproduces the desired PMNS matrix and makes predictions on the charged lepton mass matrix in the form of $\mathcal{M}_l {\mathcal{M}_l}^\dagger$. The singlet scalars play a crucial role in the process $\mu \rightarrow e\, \gamma$ in the EW-scale $\nu_R$ model as shown in \cite{Hung:2007ez} and updated below in light the aforementioned developments. The results presented in this paper contain a deep correlation between the branching ratio $B(\mu \rightarrow e\, \gamma)$ and the neutrino sector in the form of the PMNS matrix for both normal and inverted hierarchies, as well as the form of the mirror lepton mixing matrix. It will be shown that the exclusion zones in the plots of the branching ratio of 
$B(\mu \rightarrow e\, \gamma)$ versus the Yukawa coupling strengths to the singlets depend a bit on how strong the $A_4$-triplet scalars couple to the leptons.

This paper is organized as follows. First, in section II, we summarize the essence of the EW-scale $\nu_R$ model (original \cite{Hung:2006ap} and extended \cite{Hoang:2014pda}). Next, in section III, we briefly review constraints 
from electroweak precision measurements for the original model and from Higgs physics for the extended model.
In section IV, we briefly review the results of neutrino and charged lepton masses and mixings \cite{Hung:2015nva}. 
We then proceed with the actual calculations of the process $l_i \rightarrow l_j \gamma$, 
the anomalous magnetic dipole moment $\Delta a_{l_i}$ and the electric dipole moment $d_{l_i}$ for the lepton $l_i$ 
in section V. Detailed numerical analysis will be presented in section VI.  
Implications of our results concerning the possible detection of mirror leptons at the LHC and the ILC are discussed in section VII. We finally summarize and conclude in section VIII. 
A few useful formulas are collected in an Appendix.


\section{Review of  the EW-scale $\nu_R$ model}

For the sake of clarity, we review in this section the original EW-scale $\nu_R$ model \cite{Hung:2006ap} and its extended version \cite{Hoang:2014pda}.

\begin{itemize}

\item {\bf Gauge group}:

\be
SU(3)_C \times SU(2) \times U(1)_Y
\ee

There are many differences between the EW-scale $\nu_R$ model and the popular Left-Right symmetric model \cite{LR}. The first difference lies in the gauge structure of the two models: $SU(3)_C \times SU(2) \times U(1)_Y$ for the EW-scale $\nu_R$ model and $SU(3)_C \times SU(2)_L \times SU(2)_R \times U(1)_{B-L}$ for the Left-Right model. 

\item {\bf Lepton and quark $SU(2)$ doublets} (the superscript $M$ refer to mirror fermions):

SM: $l_L = \left(
	  \begin{array}{c}
	   \nu_L \\
	   e_L \\
	  \end{array}
	 \right)$; Mirror: $l_R^M = \left(
	  \begin{array}{c}
	   \nu_R^M \\
	   e_R^M \\
	  \end{array}
	 \right)$. \
	 
SM: $q_L = \left(
	  	 \begin{array}{c}
	   	  u_L \\
	     	  d_L \\
	  	\end{array}
	 	\right)$; Mirror: $q_R^M = \left(
	  	 \begin{array}{c}
	   	  u_R^M \\
	     	  d_R^M \\
	  	\end{array}
	 	\right)$.

\item {\bf Lepton and quark $SU(2)$ singlets}:

SM: $e_R; \ u_R, \ d_R$; Mirror: $e_L^M; \ u_L^M, \ d_L^M$.
		
\item {\bf Doublet Higgs fields}:

As explained in \cite{Hung:2006ap}, a Higgs doublet is needed to give masses to all {\em charged} fermions. The analysis of the properties of the 125-GeV SM-like scalar necessitates the introduction of one extra Higgs doublet as explained in 	\cite{Hoang:2014pda}. Each Higgs doublet couples to a different sector: 	$\Phi_2=(\phi_{2}^+ , \phi_{2} ^0 ) $ to the SM fermions and $\Phi_{2M}=(\phi_{2M}^+ , \phi_{2M} ^0 )$ to the mirror fermions. They develop the following vacuum-expectation-values (VEV): $\langle \Phi_2 \rangle=(0,v_2/\sqrt{2})^T$ and
$\langle \Phi_{2M} \rangle=(0,v_{2M}/\sqrt{2})^T$.

\item {\bf Triplet Higgs fields}:	 

The $SU(2)$-triplet Higgs fields form the cornerstone of the EW-scale $\nu_R$ model. As shown in \cite{Hung:2006ap}, the VEV of the $Y/2=1$ triplet gives an electroweak-scale Majorana mass to the right-handed neutrinos and the $Y/2=0$ triplet is needed to preserve the custodial symmetry so that the $\rho$ parameter equals unity at tree level. 
This is summarized below. Here we just write down the triplet Higgs fields and their VEVs. 

\begin{itemize}
			\item $\widetilde{\chi} \ (Y/2 = 1)  = \frac{1}{\sqrt{2}} \ \vec{\tau} . \vec{\chi} = 
	  \left(
	  \begin{array}{cc}
	    \frac{1}{\sqrt{2}} \chi^+ & \chi^{++} \\
	    \chi^0 & - \frac{1}{\sqrt{2}} \chi^+\\
	   \end{array}
		  \right)$ with $\langle \chi^0 \rangle = v_M$.
			\item $\xi \ (Y/2 = 0)= (\xi^+, \xi^0, \xi^-)$ (in order to restore Custodial Symmetry) with $\langle \xi^0 \rangle = v_M$.
			
	\item VEVs:
	
	$v_{2}^2 +v_{2M}^2 +  8 v_{M}^2= v^2 \approx (246 \gev)^2$.	
		\end{itemize}	

\item {\bf Singlet Higgs fields}:

The original model which is basically concerned with the energy scales which enter the seesaw mechanism contains only one $SU(2) \times U(1)_Y$-singlet Higgs field $\phi_S$ whose VEV $\langle \phi_S \rangle = v_S$ gives the Dirac mass to the neutrinos (to be summarized below). It was almost a ``toy model" in that it did not discuss lepton mixings and, in particular, the PMNS matrix $U_{\rm PMNS}$. This problem has been recently investigated by \cite{Hung:2015nva} within the framework of an $A_4$ symmetry which is applied to the neutrino sector of the EW-scale $\nu_R$ model. The upshot of this study was the introduction of an $A_4$ singlet $\phi_{0S}$ and an $A_4$-triplet \{$\phi_{iS}$\} ($i=1,2,3$). They obtain the following VEVs $v_0$ and $v_{i}$ respectively. We will summarize below the main points concerning this singlet scalar sector in the construction of $U_{\rm PMNS}$ and its implication to rare progresses such as $\mu \rightarrow e \gamma$.

\item {\bf Dirac neutrino mass}

For simplicity, we will denote the right-handed neutrino fields by $\nu_R$ from hereon.

The original model contains one singlet scalar whose VEV provides a Dirac mass for the neutrinos. A generic Yukawa coupling is of the form
\bea
\label{dirac}
{\mathcal L}_S &=& - g_{Sl} \,\bar{l}_L \ \phi_S \ l_R^M + {\rm H.c.}\\  \nonumber
         &= & - g_{Sl} (\bar{\nu}_L \ \nu_R \ + \bar{e}_L \ e_R^M) \ \phi_S + {\rm H.c.} 
\eea
With $\langle \phi_S \rangle = v_S$, one obtains the Dirac mass $m_\nu^D = g_{Sl} \ v_S $.	 

\item {\bf Majorana neutrino mass}

This is the main point of \cite{Hung:2006ap}. The electroweak-scale Majorana mass for the right-handed neutrinos is  obtained from the following coupling
\bea
\label{majorana}
L_M &= & g_M \, l^{M,T}_R \ \sigma_2 \ \tau_2 \ \tilde{\chi} \ l^M_R \\ \nonumber
&= &g_M \ \nu_R^T \ \sigma_2 \ \nu_R \ \chi^0 - \dfrac{1}{\sqrt{2}} \ \nu_R^T \ \sigma_2 \ e_R^M \ \chi^+ \\ \nonumber
&&- \dfrac{1}{\sqrt{2}} \ e_R^{M,T} \ \sigma_2 \ \nu_R \ \chi^+ + e_R^{M,T} \ \sigma_2 \ e_R^M \ \chi^{++} \,.
\eea
With $\langle \chi^0 \rangle = v_M$, the Majorana mass is given by $ M_R = g_M v_M $.

\end{itemize}


\section{Review of results of the EW-scale $\nu_R$ model as discussed in \cite{Hoang:2013jfa} and \cite{Hoang:2014pda}}

In this review section, we will discuss two sets of results for the EW-scale $\nu_R$ model obtained in \cite{Hoang:2013jfa} (the electroweak precision constraints) and \cite{Hoang:2014pda} (constraints from the 125-GeV SM-like scalar). 


\subsection{Electroweak precision constraints on the EW-scale $\nu_R$ model \cite{Hoang:2013jfa}}

The presence of mirror quark and lepton $SU(2)$-doublets can, by themselves, seriously affect the constraints coming from electroweak precision data. As noticed in \cite{Hoang:2013jfa}, the positive contribution to the S-parameter coming from the extra right-handed mirror quark and lepton doublets could be partially cancelled by the negative contribution coming from the triplet Higgs fields. Ref.~\cite{Hoang:2013jfa} has carried out a detailed analysis of the electroweak precision parameters S and T and found that there is a large parameter space in the model which satisfies the present constraints and that there is {\em no fine tuning} due to the large size of the allowed parameter space. It is beyond the scope of the paper to show more details here but a representative plot would be helpful. Fig. 1 shows the contribution of the scalar sector versus that of the mirror fermions to the S-parameter within 1$\sigma$ and 2$\sigma$.
\begin{figure}[hbtp!]
\centering
\includegraphics[width=0.5\textwidth]{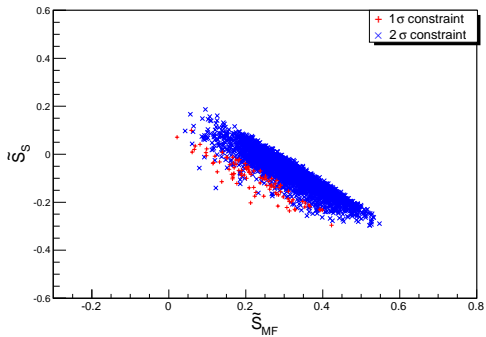} 
\caption{{\small Constrained $\tilde{S}_S$ versus $\tilde{S}_{MF}$}.}
\label{SsvsSmf}
\end{figure}
In this plot, \cite{Hoang:2013jfa} took for illustrative purpose 3500 data points that fall inside the 2$\sigma$ blue region with about 100 data points falling inside the 1$\sigma$ red region. More details can be found in \cite{Hoang:2013jfa}.


\subsection{Review of the scalar sector of the EW-scale $\nu_R$ model in light of the discovery of the 125-GeV SM-like scalar \cite{Hoang:2014pda}}

In light of the discovery of the 125-GeV SM-like scalar, it is imperative that any model beyond the SM (BSM) shows a scalar spectrum that contains at least one Higgs field with the desired properties as required by experiment. The present data from CMS and ATLAS only show signal strengths that are compatible with the SM Higgs boson. The definition of a signal strength $\mu$ is as follows 
\be\label{eq:mudef}
	\mu(H \text{-decay}) = \frac{\sigma(H \text{-decay})}{\sigma_{\rm SM}(H \text{-decay})}\,,
\ee
with
\be
\sigma(H \text{-decay}) = \sigma(H \text{-production}) \times B(H \text{-decay})\,.
\ee

To really distinguish the SM Higgs field from its impostor, it is necessary to measure the partial decay widths and the various branching ratios. In the present absence of such quantities, the best one can do is to present cases which are consistent with the experimental signal strengths. This is what was carried out in \cite{Hoang:2014pda}. 

The minimization of the potential containing the scalars shown above breaks its global symmetry $SU(2)_L \times SU(2)_R$ down to a custodial symmetry $SU(2)_D$ which guarantees at tree level $\rho = M_{W}^2/M_{Z}^2 \cos^2 \theta_W=1$ \cite{Hoang:2014pda}. The physical scalars can be grouped, based on their transformation properties under $SU(2)_D$ as follows:
	\begin{eqnarray}
		\text{five-plet (quintet)} &\rightarrow& H_5^{\pm\pm},\; H_5^\pm,\; H_5^0;\nonumber\\[0.5em]
		\text{triplet} &\rightarrow& H_{3}^\pm,\; H_{3}^0;\nonumber\\[0.5em]
		\text{triplet} &\rightarrow& H_{3M}^\pm,\; H_{3M}^0;\nonumber\\[0.5em]
		\text{three singlets} &\rightarrow& H_1^0,\; H_{1M}^0,\; H_1^{0\prime}\,.
	\end{eqnarray}
  The three custodial singlets are the CP-even states, one combination of which can be the 125-GeV scalar. In terms of the original fields, one has $H_1^0 = \phi_{2}^{0r}$,  $H_{1M}^0 = \phi_{2M}^{0r}$ and $H_1^{0\prime} = \frac{1}{\sqrt{3}} \Big(\sqrt{2}\chi^{0r}+ \xi^0\Big)$. These states mix through a mass matrix obtained from the potential and the mass eigenstates are denoted by $\widetilde{H}$, $\widetilde{H}^\prime$ and $\widetilde{H}^{\prime\prime}$, with the convention that the lightest of the three is denoted by $\widetilde{H}$, the next heavier one by $\widetilde{H}^\prime$ and the heaviest state by $\widetilde{H}^{\prime\prime}$. 
  
  To compute the signal strengths $\mu$, Ref.~\cite{Hoang:2014pda} considers $\widetilde{H} \rightarrow ZZ,~W^+W^-,~\gamma\gamma,~b\bar{b}$ and $\tau\bar{\tau}$. In addition, the cross section of $g g \rightarrow \widetilde{H}$ related to $\widetilde{H} \rightarrow g g$ was also calculated. A scan over the parameter space of the model yielded {\em two interesting scenarios} for the 125-GeV scalar: 1) {\em Dr. Jekyll}'s scenario in which $\widetilde{H} \sim H_1^0$ meaning that the SM-like component $H_1^0 = \phi_{2}^{0r}$ is {\em dominant}; 2) 
{\em Mr. Hyde}'s scenario in which $\widetilde{H} \sim H_1^{0\prime}$ meaning that the SM-like component $H_1^0 = \phi_{2}^{0r}$ is {\em subdominant}. Both scenarios give signal strengths compatible with experimental data as shown below in Fig.~(2).

\begin{figure}[hbtp!]
\label{signal2}
	\centering
	\includegraphics[width=0.5\textwidth]{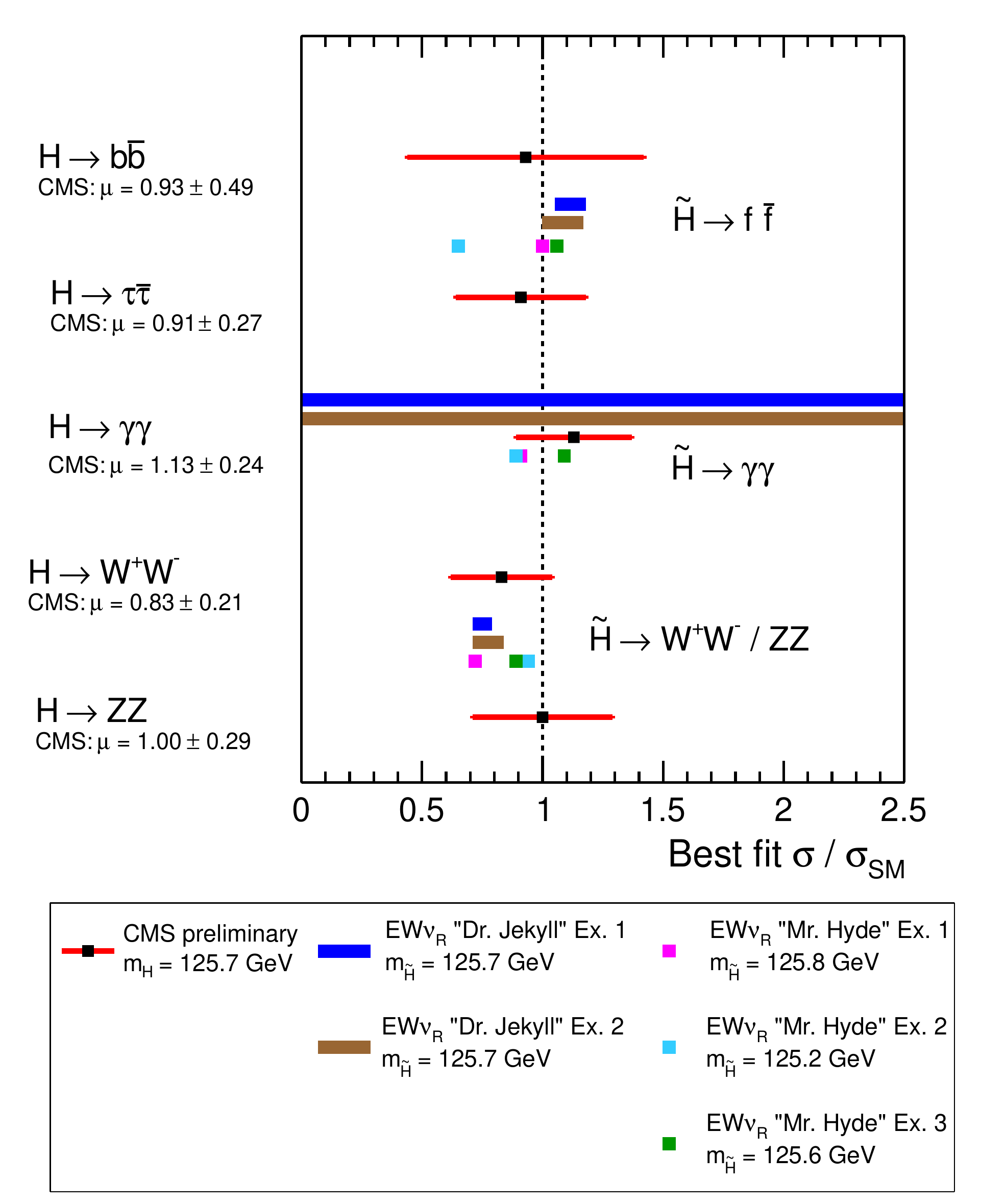}
	\caption{Predictions of signal strength $\mu(\widetilde{H} \rightarrow ~b\bar{b}, ~\tau\bar{\tau}, ~\gamma\gamma, ~W^+W^-, ~ZZ)$ in the EW-scale $\nu_R$ model for examples 1 and 2 in {\em Dr.~Jekyll} and example 1, 2 and 3 in {\em Mr.~Hyde} scenarios as discussed in \cite{Hoang:2014pda}, in comparison with corresponding best fit values by CMS \cite{h_ww_122013, h_zz_4l_122013, h_bb_102013, h_tautau_012014}.}
\end{figure}

As we can see from Fig.~(2), both SM-like scenario ({\em Dr. Jekyll}) and the {\em more interesting scenario} which is very unlike the SM ({\em Mr. Hyde}) agree with experiment. As stressed in \cite{Hoang:2014pda}, present data cannot tell whether or not the 125-GeV scalar is truly SM-like or even if it has a dominant SM-like component. It has also been stressed in \cite{Hoang:2014pda} that it is essential to measure the partial decay widths of the 125-GeV scalar to truly reveal its nature. Last but not least, in both scenarios, $H_{1M}^0 = \phi_{2M}^{0r}$ is subdominant but is essential to obtain the agreement with the data as shown in \cite{Hoang:2014pda}.

As discussed in detail in \cite{Hoang:2014pda} , for proper vacuum alignment, the potential contains a term proportional to $\lambda_5$ (Eq.~(32) of \cite{Hoang:2014pda}) and it is this term that prevents the appearance of Nambu-Goldstone (NG) bosons in the model. The would-be NG bosons acquire a mass proportional to $\lambda_5$ .

An analysis of CP-odd scalar states $H_{3}^0, H_{3M}^0 $ and the heavy CP-even states $\widetilde{H}^\prime,
\widetilde{H}^{\prime\prime}$ was presented in \cite{Hoang:2014pda}. The phenomenology of charged scalars including the doubly-charged ones was also discussed in \cite{pqaranda}.

The phenomenology of mirror quarks and leptons was briefly discussed in \cite{Hoang:2013jfa} and a detailed analysis of mirror quarks will be presented in \cite{oklahoma}. It suffices to mention here that mirror fermions decay into SM fermions through the process $q^M\rightarrow q\phi_S$, $l^M\rightarrow l\phi_S$ with $\phi_S$ ``appearing" as missing energy in the detector. Furthermore, the decay of mirror fermions into SM ones can happen outside the beam pipe and inside the silicon vertex detector. Searches for non-SM fermions do not apply in this case. It is beyond the scope of the paper to discuss these details here.


\section{Review of neutrino and charged lepton masses and mixings in the EW-scale $\nu_R$ model}

Since the ideas and notations coming out of this review will be important for the calculation of the rate of $\mu \rightarrow e \gamma$, we will present a little more details than the previous section. 

In \cite{Hung:2015nva}, a model of the Dirac part of neutrino masses was constructed using the widely popular $A_4$ symmetry. Unlike previous works on that symmetry where there was a need to introduce several (more than two and typically four or five) Higgs doublets (see the review by \cite{ma}) and where it might be very problematic with the discovery of the 125-GeV SM-like scalar, the main motivation of \cite{Hung:2015nva} is to first obtain the Cabibbo-Wolfenstein matrix \cite{CW} 
\be
\label{CW}
U_{CW} = \frac{1}{\sqrt{3}}
\left(
  \begin{array}{cccc}
  1 & 1 & 1 \\
  1 & \omega & \omega^2 \\
  1 & \omega^2 & \omega \\
  \end{array}
\right),
\ee
which is a prototype of the PMNS matrix with ``large" mixing parameters and which, upon a slight modification, could reproduce the ``experimental" $U_{\rm PMNS}$ being defined as

\be
\label{PMNS}
U_{\rm PMNS}=  U_{\nu}^{\dagger} U^{l}_{L} \,.
\ee

Under $A_4$, $(\nu,l)_L$, $(\nu, l^M)_R$, $e_R$ and $e_L^M$ transform as $\underline{3}$, where $e$ and $\nu$ are generic notations for the charged and neutral leptons.
Using the $A_4$ multiplication rule $\underline{3} \times \underline{3}= \underline{1}(11+22+33) + \underline{1}^\prime (11+\omega^2 22 + \omega 33) + \underline{1}^{\prime \prime}(11+\omega 22 + \omega^2 33)+ \underline{3} (23,31,12) + \underline{3} (32,13,21)$ with $\omega = e^{i2\pi/3}$,  it was argued in \cite{Hung:2015nva} that the appropriate set of singlet scalars is composed of an $A_4$ singlet $\phi_{0S}$ and an $A_4$-triplet \{$\phi_{iS}$\} ($i=1,2,3$). To reflect the two different ways that the $A_4$-triplet can couple to the leptons,  \cite{Hung:2015nva} wrote down the Lagrangian
\be
\label{yukawa}
{\mathcal L}_S = - \bar{l}^{0}_{L}\, (g_{0S} \phi_{0S} + g_{1S} \tilde{\phi}_S +  g_{2S} \tilde{\phi}_S )\, l^{M,0}_{R} + {\rm H.c.} \,,
\ee
where $l^{0}_L$ and $l^{M,0}_R$ are gauge eigenstates which are related to the mass eigenstates by
\be
\label{eigenstate}
l^{0}_L= U^{l}_{L} l_L \;\; , \;\;\; \;\;  l^{M,0}_{R} =  U^{l^M}_R l^{M}_{R} \,.
\ee

Using the aforementioned multiplication rule, one obtains the following matrix
\be
\label{mnu}
M_{\phi} = 
\left(
  \begin{array}{cccc}
    g_{0S}\phi_{0S} & g_{1S}\phi_{3S} & g_{2S}\phi_{2S} \\
    g_{2S}\phi_{3S} & g_{0S}\phi_{0S}  & g_{1S}\phi_{1S} \\
    g_{1S}\phi_{2S}  & g_{2S}\phi_{1S} & g_{0S}\phi_{0S}  \\
  \end{array}
\right) \, .
\ee
As shown in \cite{Hung:2015nva}, reality of neutrino Dirac masses implies that
\be
\label{real}
g_{2S} = g_{1S}^{*} \, .
\ee
Furthermore, it was shown that, with $v_0 =\langle \phi_{0S} \rangle$ and $v_i =\langle \phi_{iS} \rangle=v$, the neutrino mass matrix
\be
\label{mnu2}
M_\nu^D = 
\left(
  \begin{array}{cccc}
    g_{0S}v_0 & g_{1S}v_3 & g_{2S}v_2 \\
    g_{2S}v_3 & g_{0S}v_0 & g_{1S}v_1 \\
    g_{1S}v_2 & g_{2S}v_1 & g_{0S}v_0 \\
  \end{array}
\right) \, ,
\ee
can be diagonalized, i.e. $U_{\nu}^{\dagger} M_\nu^D U_{\nu}$, by the matrix
\be
\label{Unu}
U_{\nu}=  \frac{1}{\sqrt{3}}
\left(
  \begin{array}{cccc}
  1 & 1 & 1 \\
  1 & \omega^2 & \omega \\
  1 & \omega & \omega^2 \\
  \end{array}
\right) \, .
\ee
Notice that $U_{\nu} \equiv U^{\dagger}_{CW}$. Eqs.~(\ref{Unu}) and (\ref{mnu}) will form a basis for our subsequent discussion.

For the purpose of the subsequent sections, we rewrite Eq.~(\ref{yukawa}) as follows
\bea
{\mathcal L}_S&=& - \bar{l}_{L}\, U^{l\dagger}_{L} U_{\nu}U^{\dagger}_{\nu} M_{\phi} U_{\nu}U^{\dagger}_{\nu} U^{l^M}_R \, l^{M}_{R} +{\rm H.c.} \; \; \\
        &=& -\bar{l}_{L}\, U^{\dagger}_{\rm PMNS} \, \tilde{M}_{\phi}  \, U^{M}_{\rm PMNS} l^{M}_{R} + {\rm H.c.} \,,
\eea
where
\be
\label{Mtilde}
\tilde{M}_{\phi} = U^{\dagger}_{\nu} M_{\phi} U_{\nu} \,,
\ee
and
\be
\label{MPMNS}
U^{M}_{\rm PMNS} = U^{\dagger}_{\nu} U^{l^M}_R \,.
\ee

The above construction can be straightforwardly generalized for the right-handed leptons and left-handed mirror leptons.
Hence the total ${\mathcal L}_S$ becomes
\be
\label{totalLS}
{\mathcal L}_S = - \bar{l}_{L}\, U^{\dagger}_{\rm PMNS} \, \tilde{M}_{\phi}  \, U^{M}_{\rm PMNS} l^{M}_{R} -
\bar{l}_{R}\, U^{\prime \dagger}_{\rm PMNS} \, \tilde{M}^\prime_{\phi}  \, U^{\prime M}_{\rm PMNS} l^{M}_{L} 
+ {\rm H.c.} \,
\ee
where $\tilde{M}^\prime_{\phi} = U^\dagger_\nu M^\prime_\phi U_\nu$ and 
$M^\prime_\phi$ is the same as $M_{\phi}$ given by Eq.~(\ref{mnu}) with $g_{0S} \to g'_{0S}$,
$g_{1S} \to g'_{1S}$ and $g_{2S} \to  g'_{2S}$. Reality of the eigenvalues of $M^\prime_{\phi}$ also implies
$g'_{2S} =  g^{\prime *}_{1S}$. In analogous to $U_{\rm PMNS}$ and $U^M_{\rm PMNS}$, we have defined 
the following mixing matrices for the second term of Eq.~(\ref{totalLS}) 
\be
\label{UPMNSprime}
U^\prime_{\rm PMNS}=  U_{\nu}^{\dagger} U^{l}_{R} \,,
\ee
and
\be
\label{UPMNSMirrorprime}
U^{\prime M}_{\rm PMNS} = U^{\dagger}_{\nu} U^{l^M}_L \,,
\ee
where $U^{l}_{R}$ and $U^{l^M}_L$ are the unitary matrices relating the gauge eigenstates and the mass eigenstates
\be
\label{eigenstateright}
l^{0}_R= U^{l}_{R} l_R \;\; , \;\;\; \;\;  l^{M,0}_{L} =  U^{l^M}_L l^{M}_{L} \,.
\ee

%
\begin{figure}[hbtp!]
\centering
\includegraphics[width=0.75\textwidth]{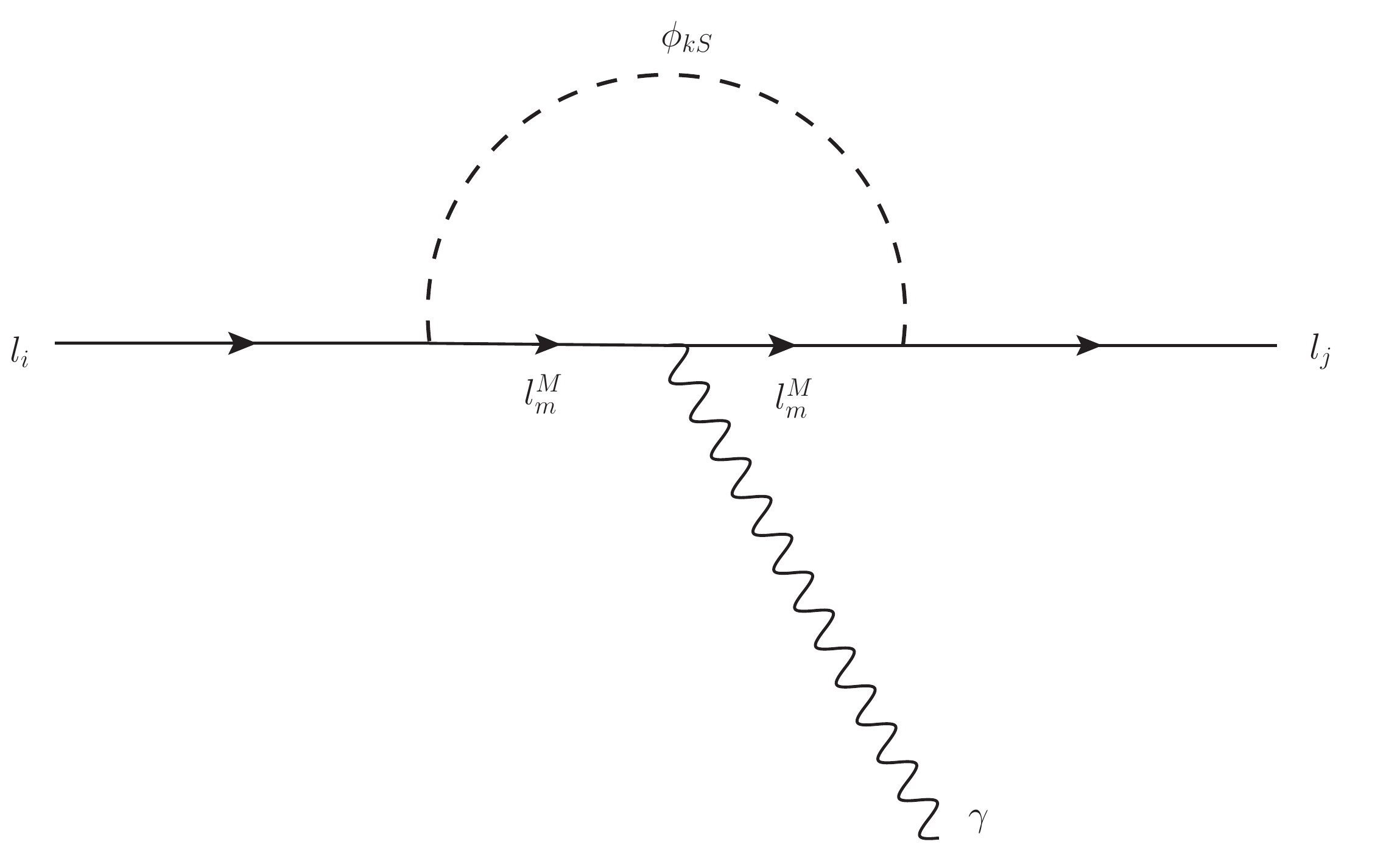} 
\caption{{\small One-loop induced Feynman diagram for $l_i \to \l_j \gamma$ in EW-scale $\nu_R$ model}.}
\label{FeynDiag}
\end{figure}

\section{The calculation}

The one-loop irreducible diagram for $l_i \to l_j \gamma$ is shown in Fig.~(\ref{FeynDiag}). 
Other two diagrams not shown are reducible associated with the one-loop dressing for the external 
fermion lines. They are crucial for  the cancellation of ultraviolet divergences and gauge invariance
in our calculation. The relevant Yukawa couplings
between the leptons, mirror leptons and the $A_4$ singlet and triplet scalars can be deduced by  
recasting the Lagrangian ${\mathcal L}_S$ in Eq.~(\ref{totalLS}) into the following component form
\bea
{\mathcal L}_{S} & = & - \sum_{k=0}^3 \sum_{i,m=1}^3  \left( \bar{l}_{Li} \, {\cal U}^{L \, k}_{im} l^M_{R m}  
+ \bar{ l}_{R i}  \, {\cal U}^{R \, k}_{im} l^M_{Lm} \right) \phi_{kS} + {\rm H.c.} 
\eea
where 
\bea
{\cal U}^{L \, k}_{im} 
& \equiv &  \left( U^\dagger_{\rm PMNS}  \cdot   M^k \cdot  U^{l^M}_{\rm PMNS} \right)_{im} \;\; , \\
& = & \sum_{j,n = 1}^3 \left( U^\dagger_{\rm PMNS} \right)_{i j}   M^k_{jn}  
\left(  U^{M}_{\rm PMNS} \right)_{nm} \; \; , 
\eea
and
\bea
{\cal U}^{R \, k}_{im} 
& \equiv & \left( U^{\prime \, \dagger}_{\rm PMNS} \cdot   M^{\prime \, k} \cdot U^{\prime \, l^M}_{\rm PMNS} \right)_{im} \; \; , \\
& = & \sum_{j,n = 1}^3 \left( U^{\prime \, \dagger}_{\rm PMNS} \right)_{i j}   M^{\prime \, k}_{jn}  
\left(  U^{\prime \, M}_{\rm PMNS} \right)_{nm} \; \; .
\eea
The matrix elements for the four matrices $M^k (k=0,1,2,3)$ are listed in Table~I.
\begin{table}
\label{M}
\caption{Matrix elements for $M^k (k=0,1,2,3)$.}
\begin{tabular}{|c|c|}
\hline
$M_{jn}^k$ & Value \\
\hline
\hline
$M^0_{12}, M^0_{13}, M^0_{21}, M^0_{23}, M^0_{31}, M^0_{32}$ & 0 \\
$M^0_{11},  M^0_{22}, M^0_{33}$ & $g_{0S}$ \\
$M^1_{11},  M^2_{11}, M^3_{11}$  & $\frac{2}{3} \mathrm{Re} \left( g_{1S} \right)$ \\ 
$M^1_{22}, M^2_{22}, M^3_{22}$  &  $\frac{2}{3} \mathrm{Re} \left( \omega^* g_{1S} \right)$  \\
$M^1_{33}, M^2_{33}, M^3_{33}$  & $\frac{2}{3} \mathrm{Re} \left( \omega g_{1S} \right)$  \\ 
$M^1_{12}, M^1_{21}$ & $\frac{2}{3} \mathrm{Re} \left( \omega g_{1S} \right)$ \\
$M^2_{12}, M^3_{21}$ & $\frac{1}{3} \left( g_{1S} + \omega g^*_{1S} \right)$ \\
$M^3_{12}, M^2_{21}$ & $\frac{1}{3} \left( g^*_{1S} + \omega^* g_{1S} \right)$ \\ 
$M^1_{13},  M^1_{31}$ & $\frac{2}{3} \mathrm{Re} \left( \omega^* g_{1S} \right)$  \\\
$M^2_{13},  M^3_{31}$ & $\frac{1}{3} \left( g_{1S} + \omega^* g^*_{1S} \right)$ \\
$M^3_{13},  M^2_{31}$ & $\frac{1}{3} \left( g^*_{1S} + \omega g_{1S} \right)$ \\
$M^1_{23}, M^1_{32}$ & $\frac{2}{3} \mathrm{Re} \left( g_{1S} \right)$ \\
$M^2_{23}, M^3_{32}$ & $\frac{2 \omega^*}{3} \mathrm{Re} \left( g_{1S} \right)$ \\
$M^3_{23}, M^2_{32}$ & $\frac{2 \omega}{3} \mathrm{Re} \left( g_{1S} \right)$  \\
\hline
\end{tabular}
\end{table}
$M^{\prime \, k}_{jn}$ can be obtained from $M^{k}_{jn}$ listed in Table~I with the following substitutions
$g_{0S} \to g^\prime_{0S}$ and  $g_{1S} \to g^\prime_{1S}$.


\subsection{The process $l_i \to l_j  \gamma \; (i \not= j)$ in EW-scale $\nu_R$ Model}

Lorentz and gauge invariance dictate the form of the amplitude 
for the process $l^-_i (p) \to l^-_j (p') + \gamma (q) $ to be
\be
{\cal M}\left( l^-_i \to l^-_j \gamma \right) = \epsilon^*_\mu(q) \bar u_j (p')  \left\{ 
i \sigma^{\mu \nu} q_\nu \left[ C^{ij}_L P_L + C^{ij}_R P_R \right] \right\} u_i(p) \; \; ,
\ee
where $P_{L,R} = (1 \mp \gamma_5 )/2$. The coefficients $C^{ij}_{L,R}$ can be extracted 
from the one-loop diagram (Fig.~(\ref{FeynDiag})),
\bea
\label{CL}
C^{ij}_L & = & + \frac{e}{16 \pi^2} \sum_{k=0}^3 \sum_{m=1}^3 
\left\{
\frac{1}{m^2_{l^M_m}}
\left[
m_i \mathcal{U}^{R \, k}_{jm} \left( \mathcal{U}^{R \, k}_{im} \right)^* 
+ m_j \mathcal{U}^{L\, k}_{jm} \left( \mathcal{U}^{L \, k}_{im} \right)^* 
\right]
{\cal I} \left( \frac{m^2_{\phi_{kS}}}{m^2_{{l^M_m}}} \right) \right. \nonumber \\ 
&\;& \;\;\;\;\;\;\;\;\;\; \;\;\;\;\;\;\; \;\;\;\;\;\;\;+ \left. \frac{1}{m_{l^M_m}} \mathcal{U}^{R \, k} _{jm} \left( \mathcal{U}^{L\, k}_{im} \right)^* {\cal J} \left( \frac{m^2_{\phi_{kS}}}{m^2_{{l^M_m}}} \right)
\right\} \;\; ,\\
\label{CR}
C^{ij}_R & = & + \frac{e}{16 \pi^2} \sum_{k=0}^3 \sum_{m=1}^3 
\left\{
\frac{1}{m^2_{l^M_m}}
\left[
m_i \mathcal{U}^{L\, k}_{jm} \left( \mathcal{U}^{L\, k}_{im} \right)^* 
+ m_j \mathcal{U}^{R \, k}_{jm} \left( \mathcal{U}^{R \, k}_{im} \right)^* 
\right]
{\cal I} \left( \frac{m^2_{\phi_{kS}}}{m^2_{{l^M_m}}} \right) \right. \nonumber \\ 
&\;& \;\;\;\;\;\;\;\;\;\; \;\;\;\;\;\;\; \;\;\;\;\;\;\;+ \left. \frac{1}{m_{l^M_m}} \mathcal{U}^{L\, k}_{jm} 
\left( \mathcal{U}^{R\, k}_{im} \right)^* {\cal J} \left( \frac{m^2_{\phi_{kS}}}{m^2_{{l^M_m}}} \right)
\right\} \; \; .
\eea
Here we have assumed the mirror lepton masses are much larger than the external fermion masses 
$m_{l^M_m} \gg m_{i,j}$ and set $m_{i,j} \to 0$ in the loop functions 
${\cal I}(r)$ and ${\cal J}(r)$, which are simply given by
\bea
\label{I}
{\cal I}(r) & = & \frac{1}{12 (1 - r)^4} \left[ - 6 r^2 \log r + r ( 2 r^2 + 3 r - 6 ) + 1 \right] \;\; , \\
\label{J}
{\cal J}(r) & = & \frac{1}{2 (1 - r)^3} \left[ - 2 r^2 \log r + r ( 3 r - 4) + 1 \right] \; \; .
\eea
In our numerical work for $\mu \to e \gamma$ presented in section VI, 
we will consider the mirror lepton masses of the order 
a few hundred GeV and the $A_4$ singlet and triplet scalar masses of the order 10 MeV, thus the ratio 
$r=m^2_{\phi_{kS}}/m^2_{{l^M_m}} \sim 10^{-8}$ is very tiny. For all practical purposes, one can replace 
Eqs.~(\ref{I}) and (\ref{J}) by the limits 
$\lim_{r\to 0} {\mathcal I}(r) = 1/12$ and  $\lim_{r\to 0} {\mathcal J}(r) = 1/2$ respectively. Formulas
of $\cal I$ and $\cal J$ for the general case of $m_{i,j} \neq 0$ are given in the Appendix.

The partial width for $l_i \to l_j \gamma$ is given by
\be
\Gamma \left( l_i \to l_j \gamma \right) = \frac{1}{16 \pi} m^3_{l_i} \left( 1 - \frac{m^2_{l_j}}{m^2_{l_i}} \right)^3 
\left( \vert C^{ij}_L \vert^2 + \vert C^{ij}_R \vert^2 \right) \; \; .
\ee


\subsection{Magnetic Dipole Moment}

The magnetic dipole moment anomaly for lepton $l_i$ can be easily extracted from the above calculation 
with the following result
\bea
\Delta a_{l_i} & = & \frac{2 m_{l_i}}{e} \left( \frac{C^{ii}_L + C^{ii}_R}{2} \right) \nonumber \\
& = & + \frac{1}{16 \pi^2} 
\left\{ 
\sum_{k=0}^3 \sum_{m=1}^3  2 \left( \vert \mathcal{U}^{L\, k}_{im} \vert^2 
+ \vert \mathcal{U}^{R\, k}_{im} \vert^2 \right)  
\frac{m^2_{l_i}}{m^2_{{l^M_m}}}
{\cal I} \left( \frac{m^2_{\phi_{kS}}}{m^2_{{l^M_m}}} \right) \right. \nonumber \\
&& \;\;\;\;\;\;\;\; + \left. 
 \sum_{k=0}^3 \sum_{m=1}^3 \, \mathrm{Re} \left(  \mathcal{U}^{L \, k}_{im} 
\left( \mathcal{U}^{R\, k}_{im} \right)^* \right)  
\frac{m_{l_i}}{m_{{l^M_m}}}
{\cal J} \left( \frac{m^2_{\phi_{kS}}}{m^2_{{l^M_m}}} \right)
\right\}  \;\; .
\label{mdm}
\eea


\subsection{Electric Dipole Moment}

The electric dipole moment operator for a fermion $f$ is usually defined as
\be
{\cal L}_{\rm EDM} = - i \frac{d_f}{2} \bar f \sigma^{\mu \nu} \gamma_5 f F_{\mu \nu} \; \; ,
\ee
where $F_{\mu\nu}$ is the electromagnetic field strength and the coefficient $d_f$ the electric dipole moment. 
The electric dipole moment for lepton $l_i$ can also be easily extracted from the above calculation with the result
\bea
d_{l_i} & = & \frac{i}{2} \left( C^{ii}_L - C^{ii}_R \right) \;\; , \nonumber \\
&= & + \frac{e}{16 \pi^2} 
\sum_{k=0}^3 \sum_{m=1}^3 \frac{1}{m_{{l^M_m}}} \mathrm{Im} \left(  \mathcal{U}^{L\, k}_{im} 
\left( \mathcal{U}^{R\, k}_{im} \right)^* \right) 
{\cal J} \left( \frac{m^2_{\phi_{kS}}}{m^2_{{l^M_m}}} \right)  \;\; .
\label{edm2}
\eea
 

\section{Numerical Analysis}

The branching ratio ${\rm B}(\mu \to e \gamma)$ is given by
\be
{\rm B}(\mu \to e \gamma) = \tau_\mu \cdot \Gamma \left( l_i \to l_j \gamma \right)
\ee
where $\tau_\mu$ is the lifetime of the muon \cite{PDG2014}
\be
\tau_\mu = ( 2.1969811 \pm 0.0000022 ) \times 10^{-6} \; {\rm s} \;\; .
\ee  

In our numerical analysis, we will adopt the following approach:

\begin{itemize}
\item
For the masses of the singlet scalars $\phi_{kS}$, we take
$$
m_{\phi_{0S}} : m_{\phi_{1S}} : m_{\phi_{2S}} : m_{\phi_{3S}}= M_S : 2 M_S : 3 M_S : 4  M_S
$$
with a fixed common mass $M_S = 10$ MeV. As long as $m_{\phi_{kS}} \ll m_{l^M_m}$, our results will not
be affected much by the exact mass relations among these singlet scalars. 

\item
For the masses of the mirror lepton $l^M_m$, we take 
$$
m_{l^M_m} = M_{\rm mirror} + \delta_m
$$
with $\delta_1 = 0$, $\delta_2 = 10$ GeV, $\delta_3 = 20$ GeV and vary the
common mass $M_{\rm mirror}$ from 100 GeV to 800 GeV. 

\item
We assume all the Yukawa couplings $g_{0S}$, $g_{1S}$, $g_{2S}$, $g^\prime_{0S}$, $g^\prime_{1S}$, 
and $g^\prime_{2S}$ 
to be all real\footnote{In this study, we do not analyze the possibility of electric dipole moments 
for the charged leptons in which complex Yukawa couplings must be assumed.}.
As mentioned before, $g_{2S} = ( g_{1S} )^*$ and $g^\prime_{2S} = (g^{\prime}_{1S})^*$ due to the 
reality of the mass eigenvalues of the Dirac neutrino masses. For simplicity, we also take 
$g_{0S} = g^\prime_{0S}$, $g_{1S} = g^\prime_{1S}$ and study the following 6 cases:
\begin{enumerate}
\item
$g_{0S} \neq 0, \; g_{1S} = 0$. The $A_4$ triplet terms are switched off.
\item
$g_{1S} = 10^{-2} \times g_{0S}$. The $A_4$ triplet couplings are merely one percent of the singlet ones.
\item
$g_{1S} = 10^{-1} \times g_{0S}$. The $A_4$ triplet couplings are 10 percent of the singlet ones.
\item
$g_{1S} = 0.5 \times g_{0S}$. The $A_4$ triplet couplings are one half of the singlet ones.
\item
$g_{1S} = g_{0S}$. Both $A_4$ singlet and triplet terms have the same weight. 
\item
$g_{0S} = 0, \; g_{1S} \neq 0$. The $A_4$ singlet terms are switched off.
\end{enumerate}

\item
For the three unknown mixing matrices 
$U^{M}_{\rm PMNS}$, $U^\prime_{\rm PMNS}$ and $U^{\prime M}_{\rm PMNS}$, 
we will consider two scenarios:

\begin{itemize}
\item{Scenario 1}
$$U^{M}_{\rm PMNS} = U^\prime_{\rm PMNS} =U^{\prime M}_{\rm PMNS} =U^\dagger_{CW}$$

\item{Scenario 2}
$$U^{M}_{\rm PMNS} = U^\prime_{\rm PMNS} =U^{\prime M}_{\rm PMNS} =U_{\rm PMNS}$$

Recall that the standard parameterization of the PMNS matrix is given by
\bea
U_{\rm PMNS} = \left( 
\begin{array}{ccc}
c_{12}c_{13} & s_{12}c_{13} & s_{13} e^{-i\delta} \\
-s_{12}c_{23}-c_{12}s_{23}s_{13}e^{i \delta} & c_{12}c_{23}-s_{12}s_{23}s_{13}e^{i \delta} & s_{23}c_{13}\\
s_{12}s_{23}-c_{12}c_{23}s_{13}e^{i \delta} & -c_{12}s_{23}-s_{12}c_{23}s_{13}e^{i \delta} & c_{23}c_{13}
\end{array}
\right)  \cdot P \nonumber
\eea
where $s_{ij} \equiv \sin \theta_{ij}$,  $c_{ij} \equiv \cos \theta_{ij}$ and 
$P={\rm Diag}(1, e^{i \alpha_{21}/2}, e^{i \alpha_{31}/2})$  is the Majorana phase matrix.
We will ignore the Majorana phases in this analysis.

In Table~II we list the $1\sigma$ range of the mixing parameters as given by the
recent analysis of global three-neutrino oscillation data in~\cite{Capozzi:2015uma,Capozzi:2013csa}.
With the central values for the mixing parameters given in Table~II as inputs, 
we obtain two possible solutions of the PMNS matrix:
\bea
U^{\rm NH}_{\rm PMNS} =
\left(
\begin{array}{ccc}
0.8221 & 0.5484 & -0.0518 + 0.1439 i \\
-0.3879 + 0.07915 i & 0.6432 + 0.0528 i & 0.6533 \\
0.3992 + 0.08984 i & -0.5283 + 0.05993 i & 0.7415
\end{array}
\right) \nonumber
\eea
for normal hierarchy, and
\bea
U^{\rm IH}_{\rm PMNS} =
\left(
\begin{array}{ccc}
0.8218 & 0.5483 & -0.08708 + 0.1281 i \\
-0.3608 + 0.0719 i & 0.6467 + 0.04796 i & 0.6664 \\
0.4278 + 0.07869 i & -0.5254 + 0.0525 i & 0.7293
\end{array}
\right) \nonumber
\eea
for inverted hierarchy.
For each scenario, we consider these two possible solutions for the $U_{\rm PMNS}$.
Due to the small differences between these two solutions, we expect our results 
are not too sensitive to the neutrino mass hierarchies.
\end{itemize}

\item
Limits on $B(\mu \to e \gamma)$ from MEG experiment \cite{Adam:2013mnn} and its projected sensitivity \cite{Renga:2014xra}:
\bea
\label{MEGLimitCurrent}
B(\mu \to e \gamma) & \leq & 5.7 \times 10^{-13} \; {\rm (90 \, C.L.) [MEG, \, 2013]} \; ,  \\
B(\mu \to e \gamma) & \sim  & 4\times 10^{-14} \; {\rm [Projected \, Sensitivity]} \; . 
\eea

\item 
$\Delta a_\mu$ from E821 experiment \cite{muonanomaly}:

\be
\label{E821Limit}
\Delta a_\mu \equiv  a^{\rm exp}_{\mu} - a^{\rm SM}_{\mu} = 288(63)(49) \times 10^{-11} \; .
\ee

\end{itemize}

Since the dominant contributions to the loop amplitude arise from the mass insertion of the internal 
mirror lepton line in Fig.~(\ref{FeynDiag}), only the last terms in Eqs.~(\ref{CL}), (\ref{CR}) and (\ref{mdm}) are significant numerically. 
As long as $m_{\phi_{kS}} \ll m_{l^M_m}$, the current MEG limit (Eq.~(\ref{MEGLimitCurrent}))
on the branching ratio $B(\mu \to e \gamma)$ imposes the constraint
\be
\left\vert \sum_{k,m} \mathcal{U}^{R \, k} _{1m} \left( \mathcal{U}^{L\, k}_{2m} \right)^* 
\left( \frac{ 100 \, {\rm GeV} }{m_{l^M_m}} \right) \right\vert^2
+
\left\vert \sum_{k,m} \mathcal{U}^{L \, k} _{1m} \left( \mathcal{U}^{R\, k}_{2m} \right)^* 
\left( \frac{100 \, {\rm GeV}}{m_{l^M_m}} \right) \right\vert^2 
\leq 7.9 \times 10^{-19} \; , \nonumber
\ee
while the result from the Brookhaven E821 experiment on $\Delta a_\mu$ (Eq.~(\ref{E821Limit})) imposes
\be
\sum_{k,m} \mathrm{Re} \left(  \mathcal{U}^{L \, k}_{2m} 
\left( \mathcal{U}^{R\, k}_{2m} \right)^* \right)  
\left( \frac{100 \, {\rm GeV}}{m_{l^M_m}} \right) \leq 8.6 \times 10^{-4} \; . \nonumber
\ee 

\begin{table}
\label{mixingparameters}
\caption{Mixing parameters from global three-neutrino oscillation data taken from~\cite{Capozzi:2015uma,Capozzi:2013csa}.}
\begin{tabular}{|c||c|c|}
\hline
Mixing Parameters & Normal Hierarchy & Inverted Hierarchy \\
\hline
\hline
$\sin^2\theta_{12}$ & $0.308 \pm 0.017$ & $0.308 \pm 0.017$ \\
$\sin^2\theta_{23}$ & $0.437^{+0.033}_{-0.023}$ & $0.455^{+0.139}_{-0.031}$\\
$\sin^2\theta_{13}$ & $0.0234^{+0.0020}_{-0.0019}$ & $0.024^{+0.0019}_{-0.0022}$\\
$\delta/\pi$ & $1.39^{+0.38}_{-0.27}$ & $1.31^{+0.29}_{-0.33}$\\
$\delta m^2 = m_2^2 - m_1^2$ & $( 7.54^{+0.26}_{-0.22} ) \times 10^{-5} {\rm eV}^2$ &  $( 7.54^{+0.26}_{-0.22} ) \times 10^{-5} {\rm eV}^2$\\
$\Delta m^2 = \vert m_3^2 - (m_1^2 + m_2^2) /2 \vert$ & $(2.43 \pm 0.06 ) \times 10^{-3} {\rm eV}^2$ & $(2.38 \pm 0.06 ) \times 10^{-3} {\rm eV}^2$\\
\hline
\end{tabular}
\end{table}
%


\begin{figure}[hbtp!]
\centering
\includegraphics[width=0.75\linewidth]{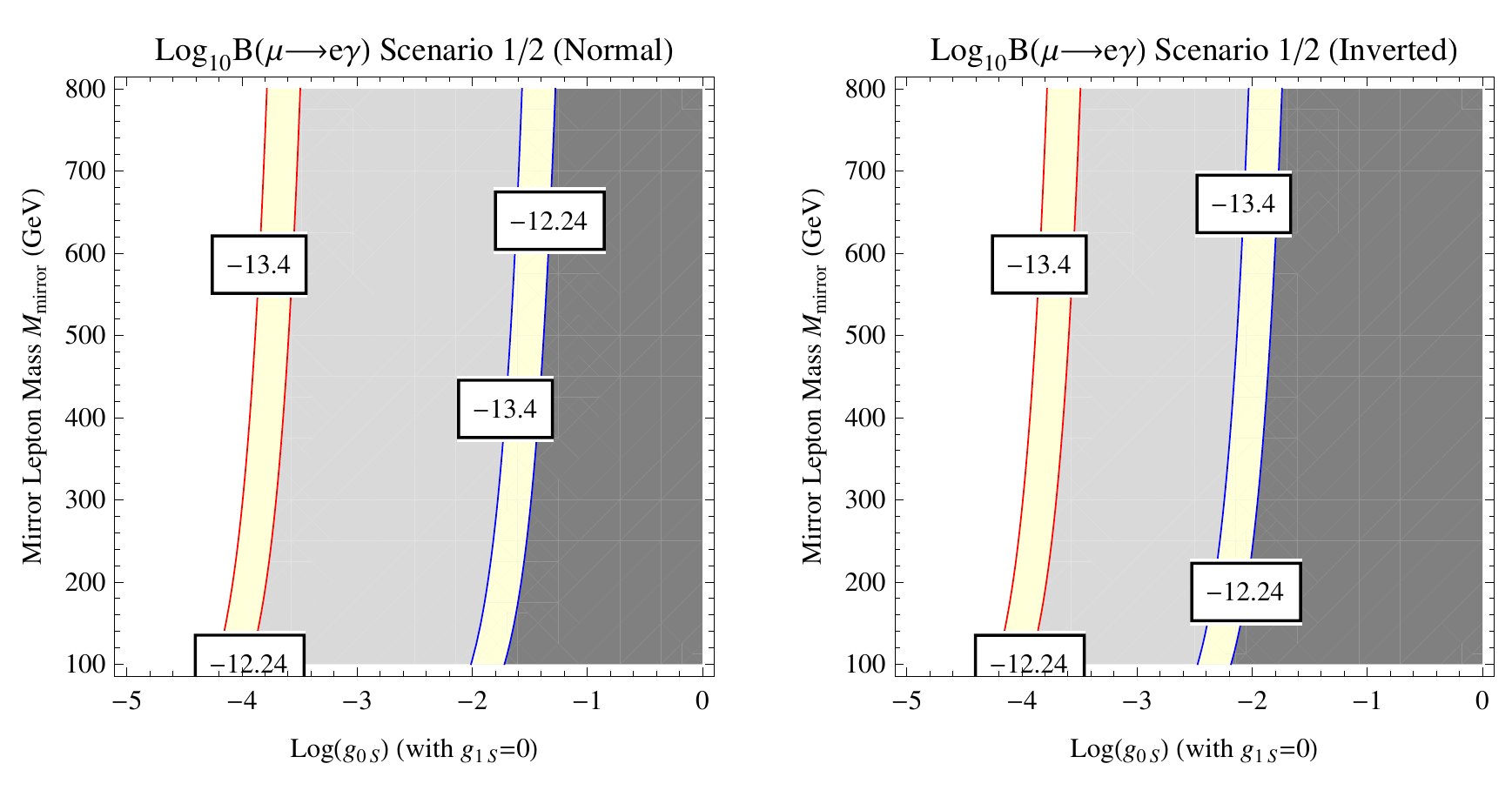} 
\includegraphics[width=0.75\linewidth]{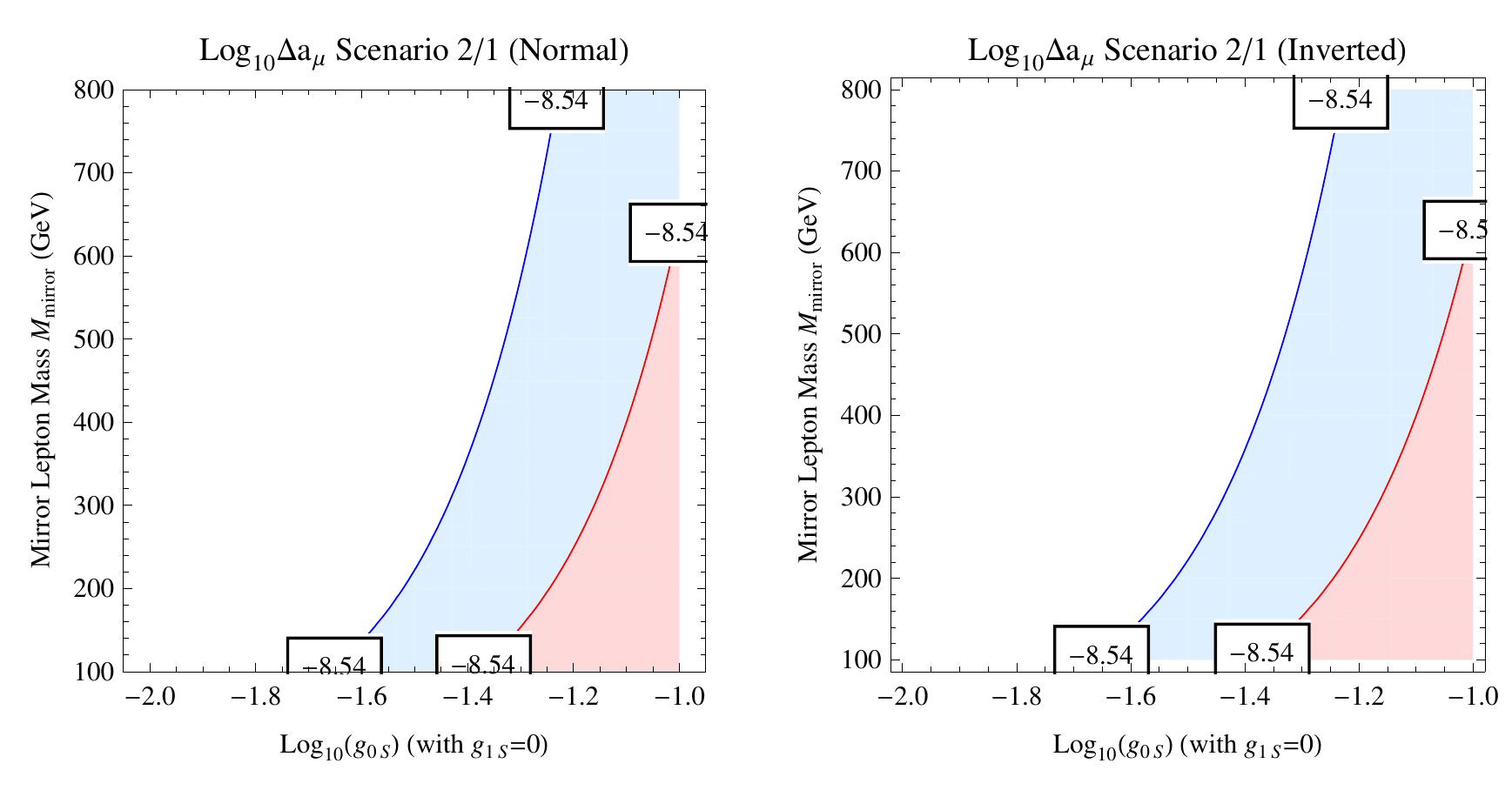} 
\caption{\small Contour plots of ${\rm Log}_{10} B(\mu \to e \gamma)$ (top panel) and 
${\rm Log}_{10} \Delta a_\mu$ (bottom panel) 
on the $(g_{0S},M_{\rm mirror})$ plane for normal (left panel) and inverted (right panel) hierarchy in scenarios 1 (red curves) 
and 2 (blue curves) with $g_{0S}=g^\prime_{0S}$ and $g_{1S}=g^\prime_{1S}=0$. 
For details of other input parameters, one can refer to the text in section VI.
} 
\label{fig4}
\end{figure}

\begin{figure}[hbtp!]
\centering
\includegraphics[width=0.75\linewidth]{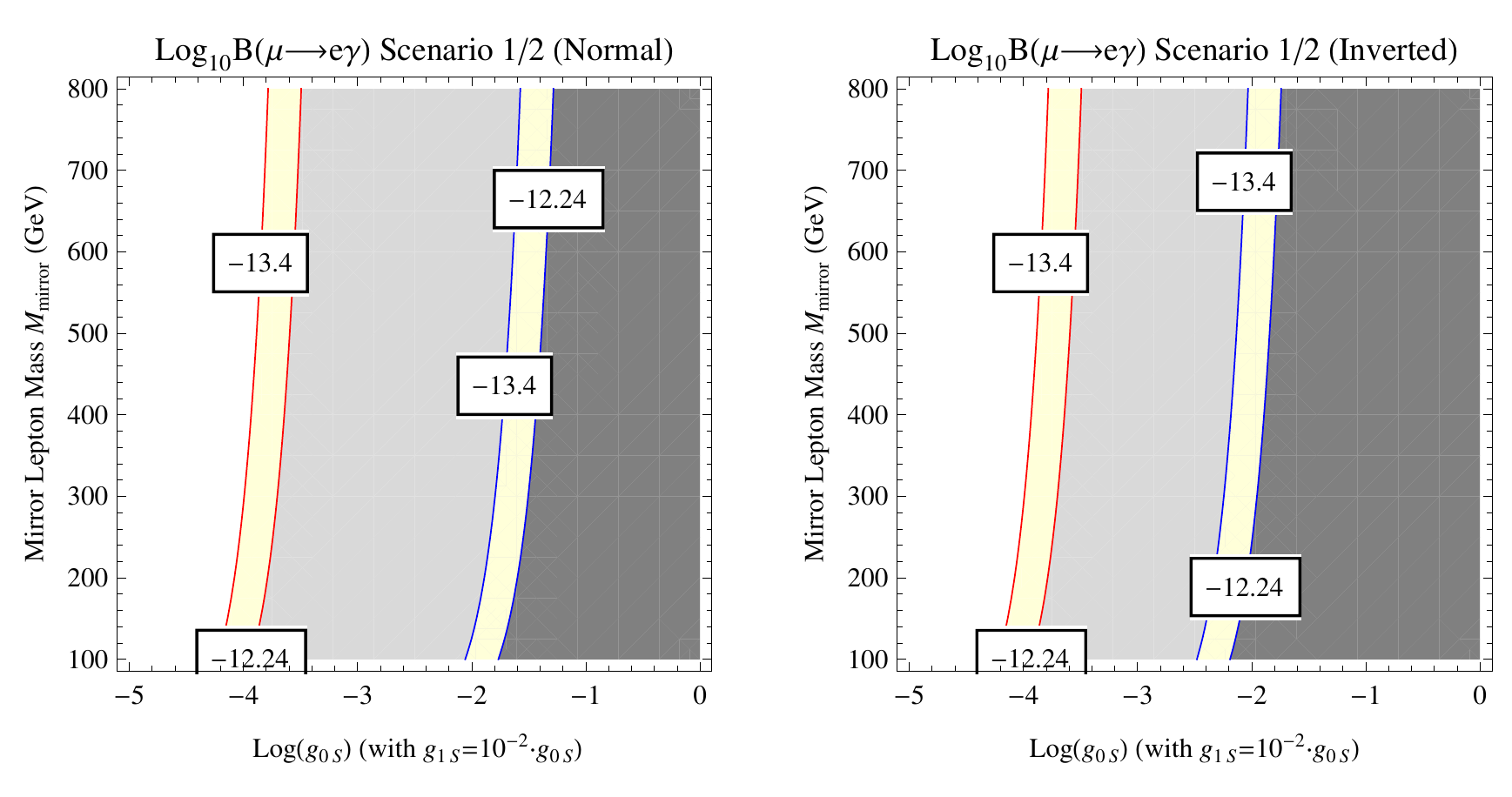} 
\includegraphics[width=0.75\linewidth]{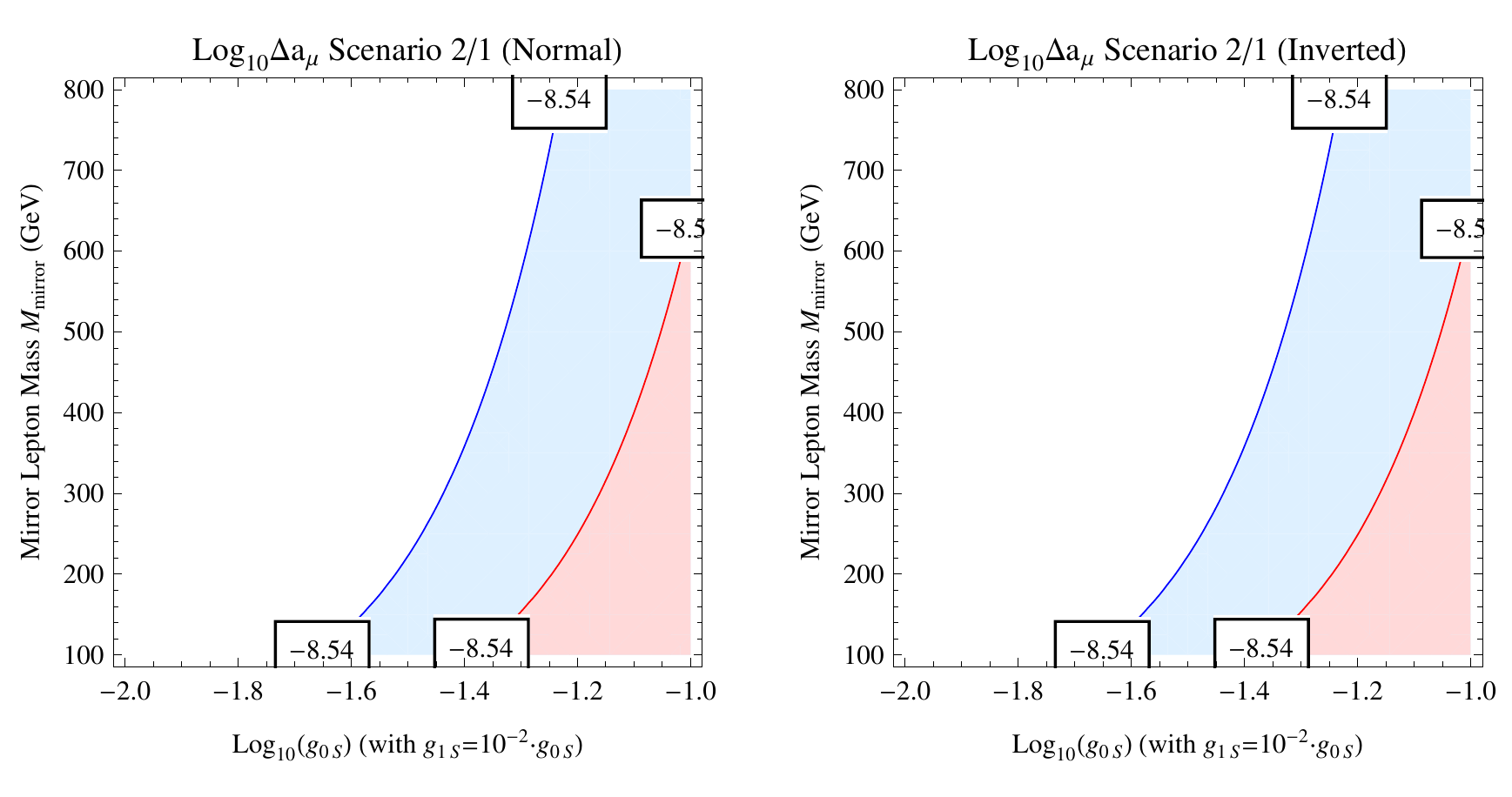} 
\caption{Same as Fig.~(\ref{fig4}) with 
$g_{0S}=g^\prime_{0S}$  and $g_{1S}=g^\prime_{1S}=10^{-2} \cdot g_{0S}$ instead.}
\label{fig5}
\end{figure}

\begin{figure}[hbtp!]
\centering
\includegraphics[width=0.75\linewidth]{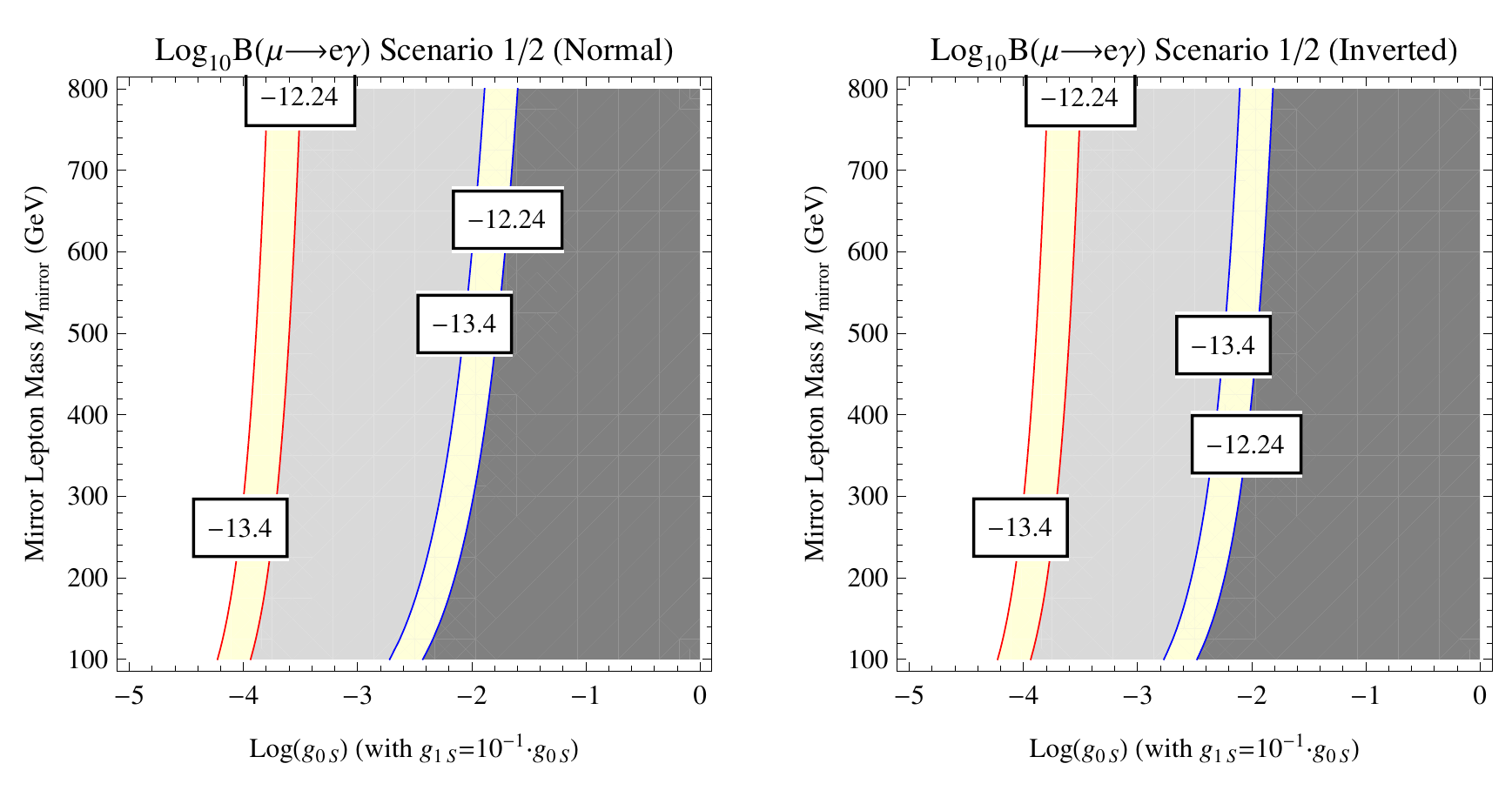} 
\includegraphics[width=0.75\linewidth]{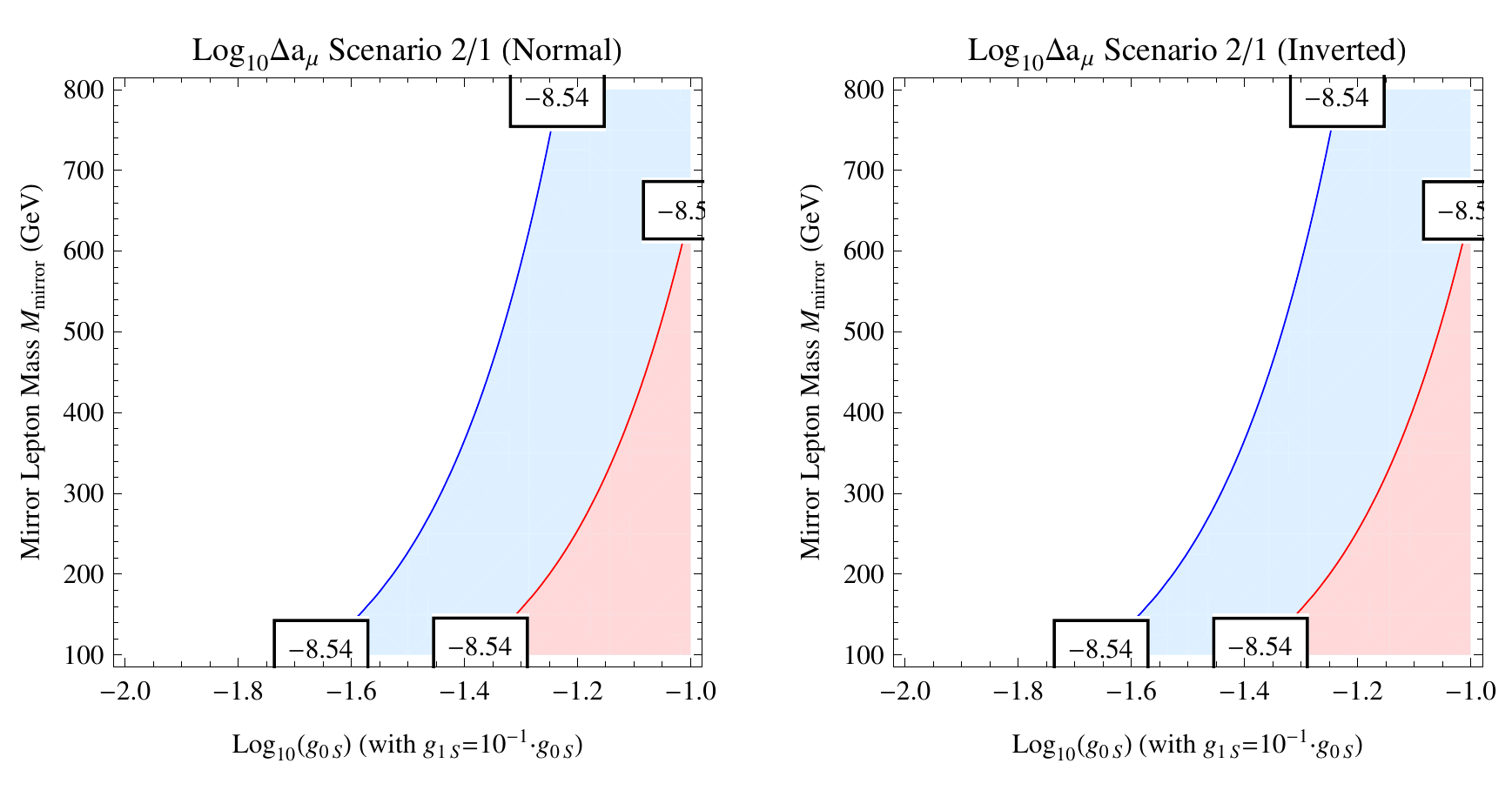} 
\caption{Same as Fig.~(\ref{fig4}) with 
$g_{0S}=g^\prime_{0S}$  and $g_{1S}=g^\prime_{1S}=10^{-1} \cdot g_{0S}$ instead.}
\label{fig6}
\end{figure}

\begin{figure}[hbtp!]
\centering
\includegraphics[width=0.75\linewidth]{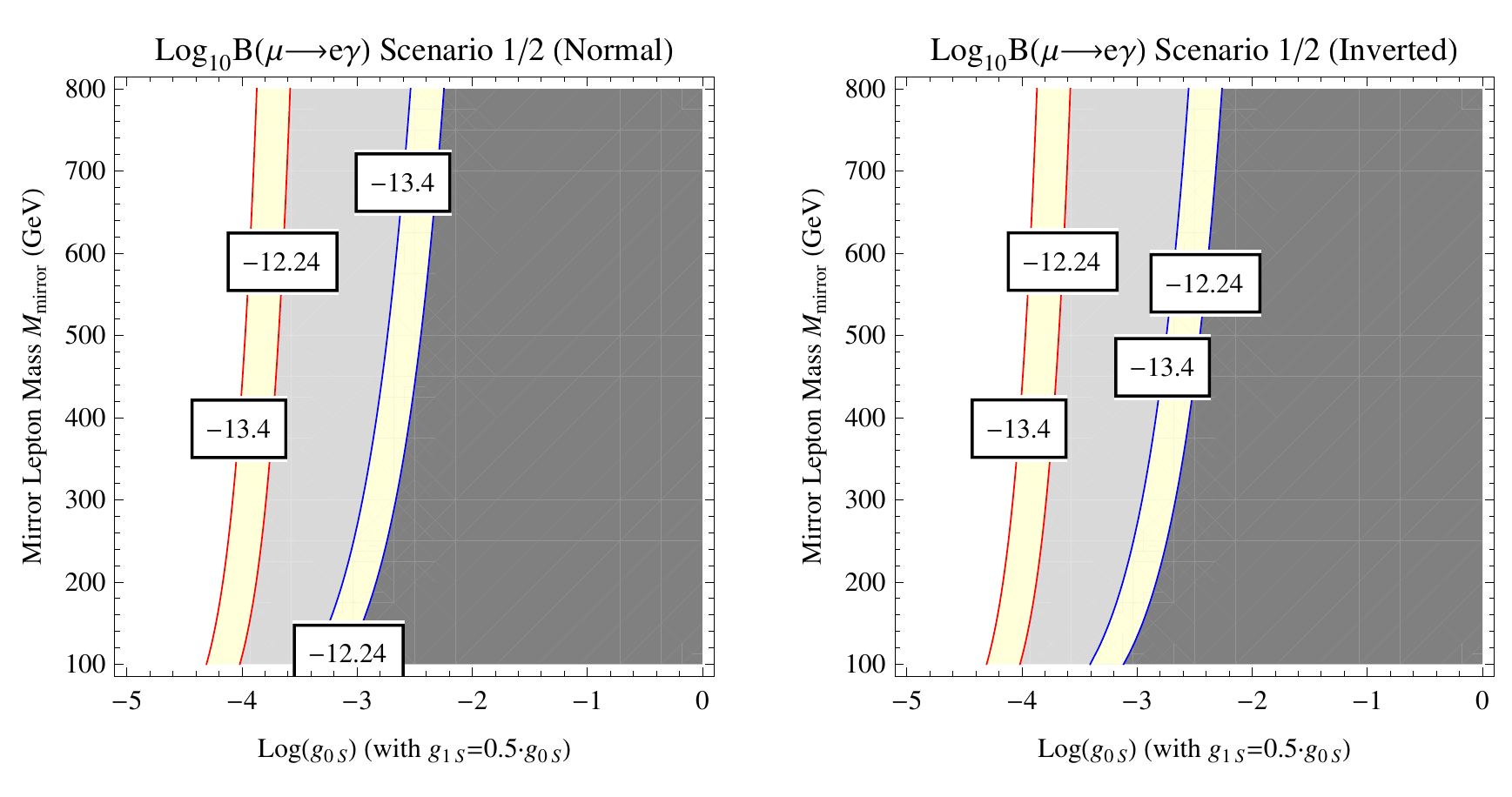} 
\includegraphics[width=0.75\linewidth]{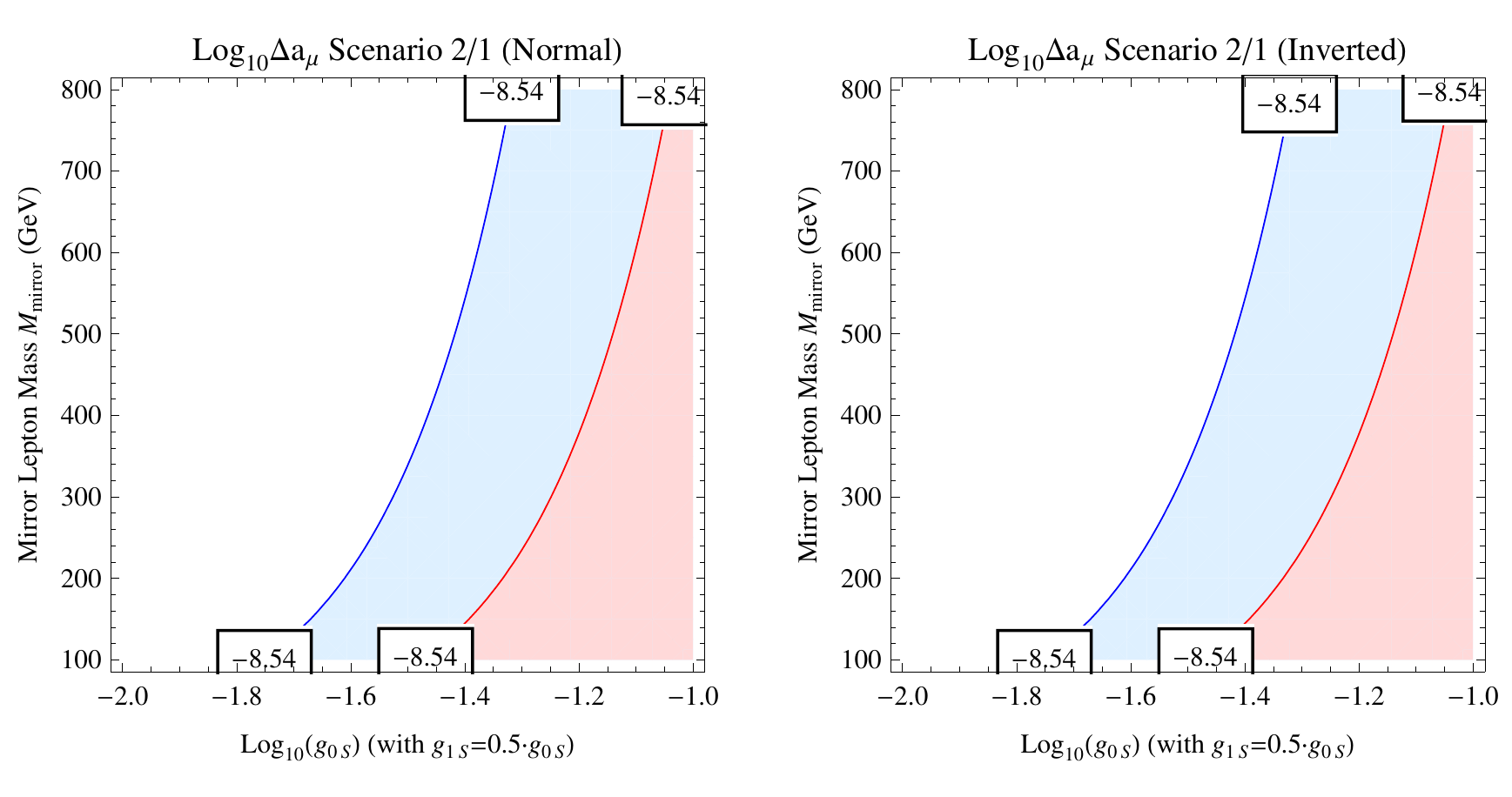} 
\caption{Same as Fig.~(\ref{fig4}) with 
$g_{0S}=g^\prime_{0S}$  and $g_{1S}=g^\prime_{1S}=0.5 \cdot g_{0S}$ instead.}
\label{fig7}
\end{figure}

\begin{figure}[hbtp!]
\centering
\includegraphics[width=0.75\linewidth]{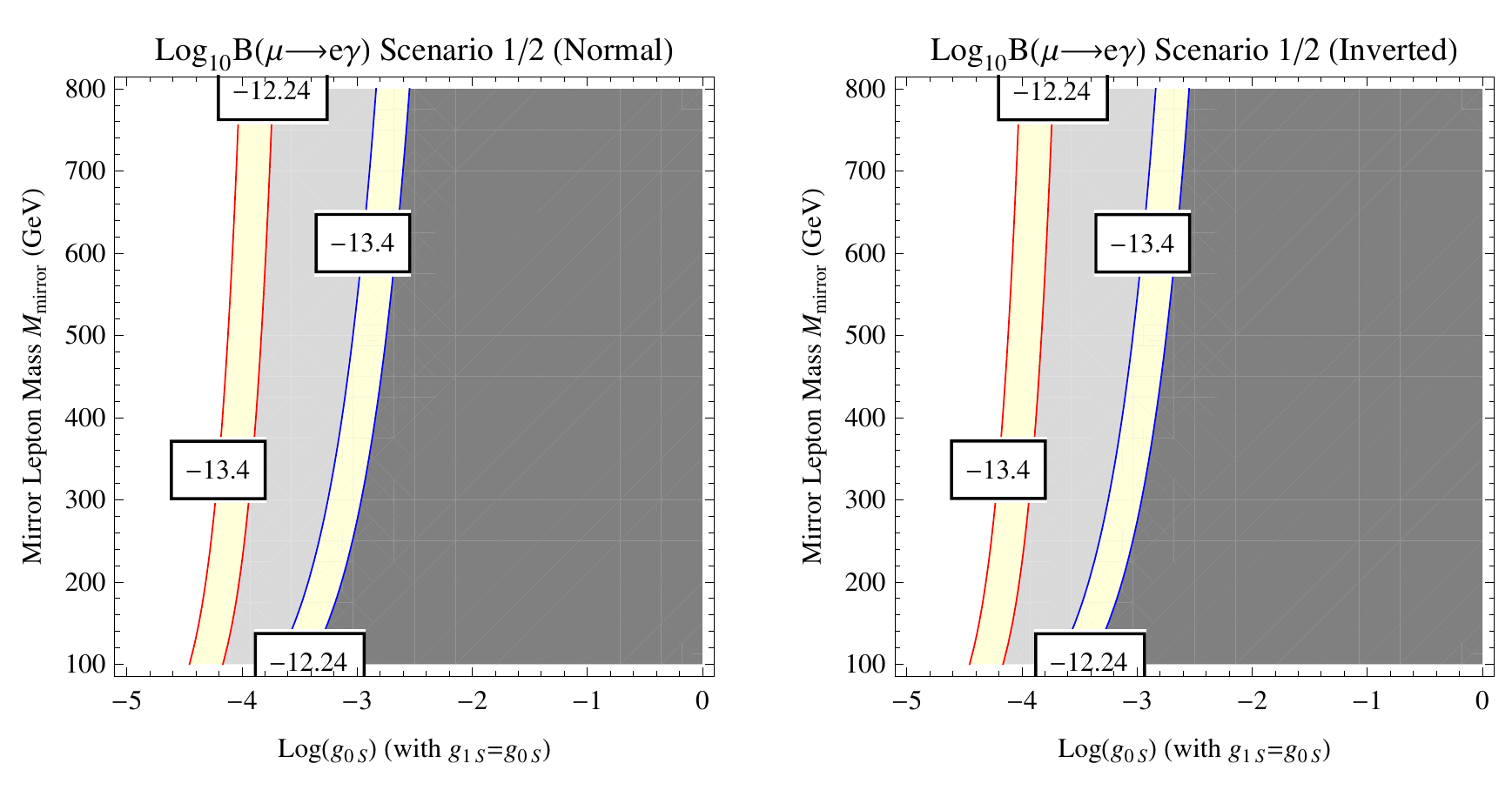} 
\includegraphics[width=0.75\linewidth]{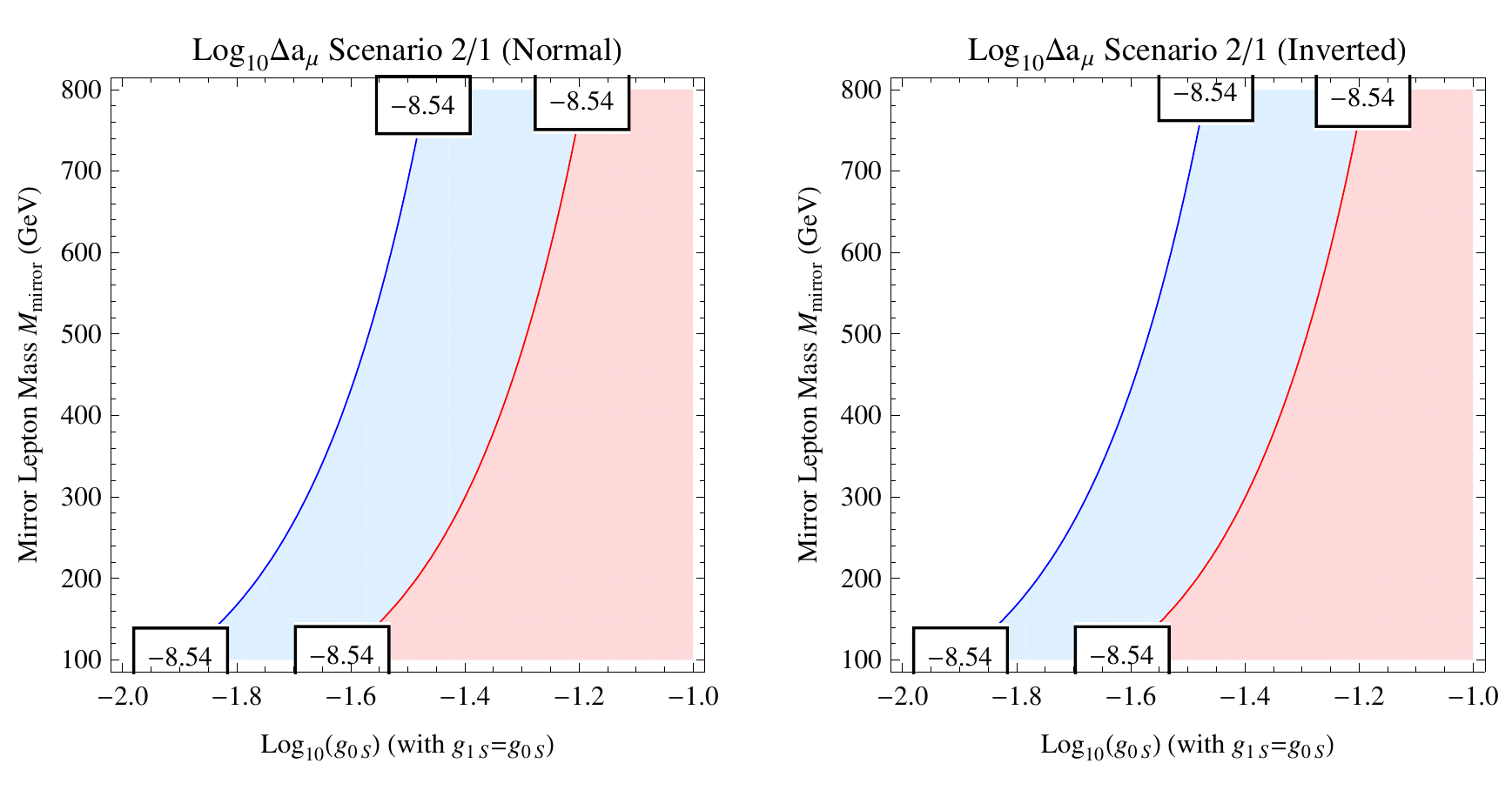} 
\caption{Same as Fig.~(\ref{fig4}) with 
$g_{0S}=g^\prime_{0S} = g_{1S}=g^\prime_{1S}$ instead.}
\label{fig8}
\end{figure}

\begin{figure}[hbtp!]
\centering
\includegraphics[width=0.75\linewidth]{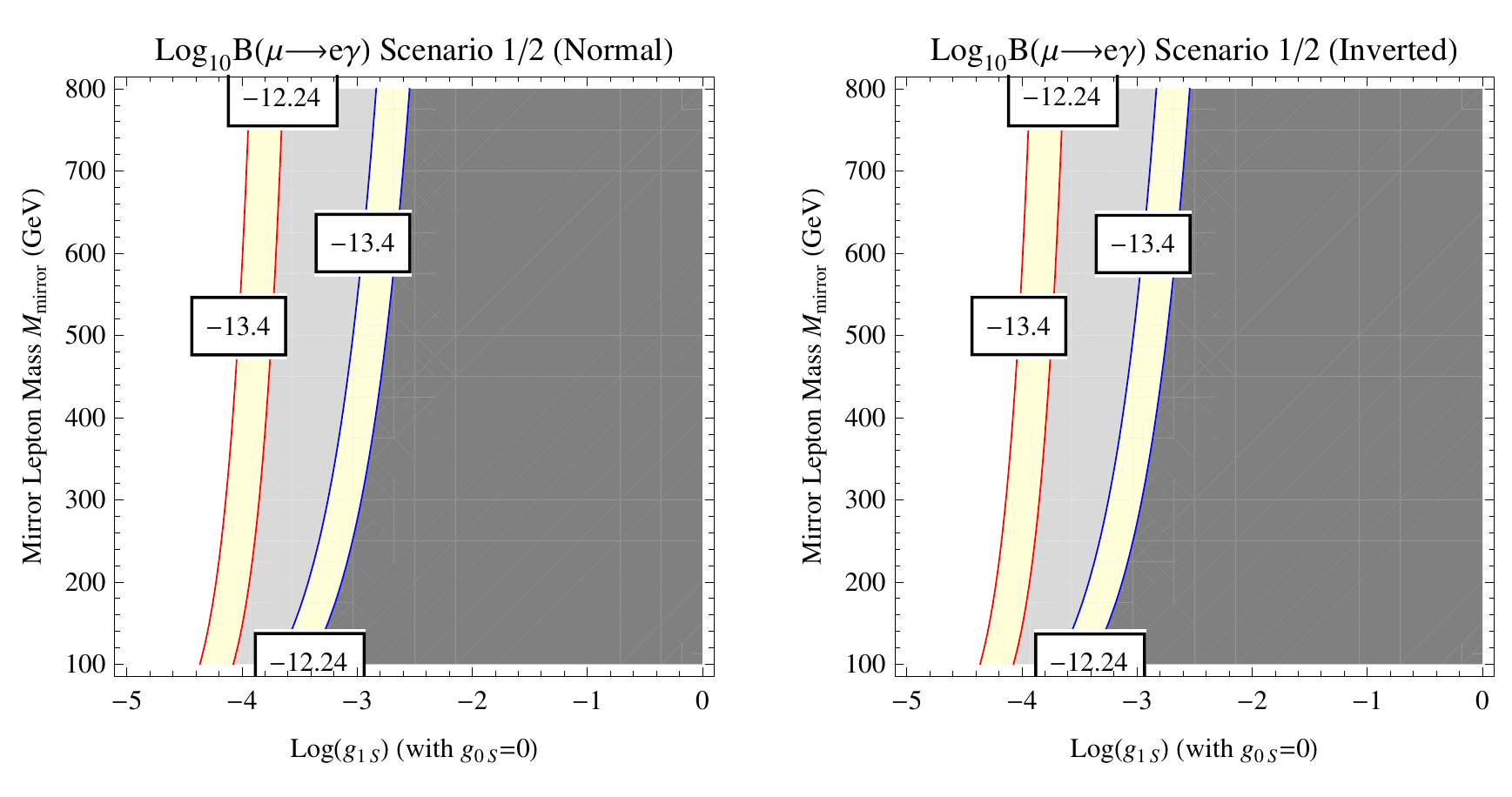} 
\includegraphics[width=0.75\linewidth]{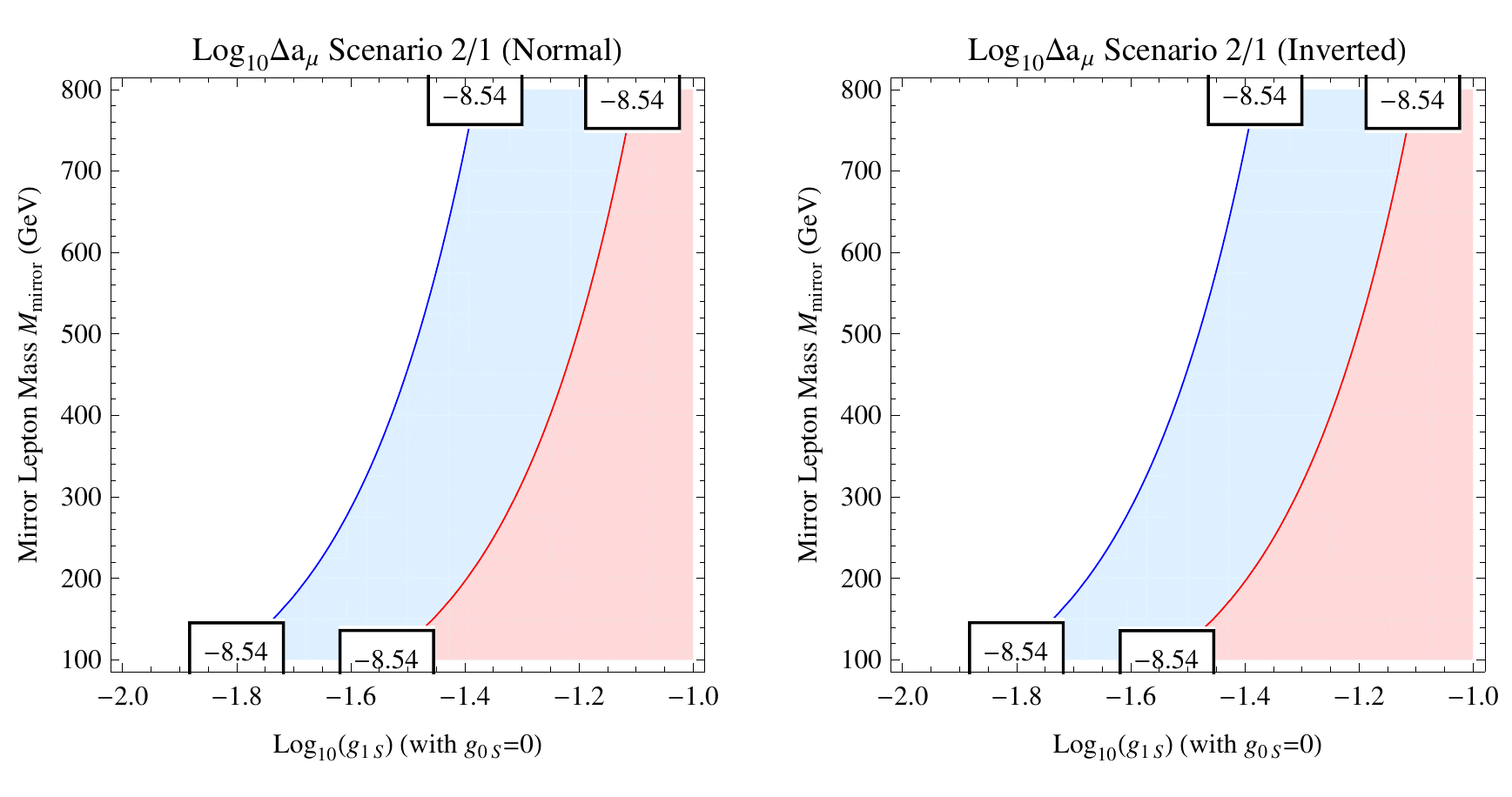} 
\caption{Same as Fig.~(\ref{fig4}) with
$g_{0S}=g^\prime_{0S}=0$ and $g_{1S}=g^\prime_{1S}$ instead.}
\label{fig9}
\end{figure}


In Figs.~(\ref{fig4})-(\ref{fig9}) we plot the contour of 
${\rm Log}_{10}B(\mu \to e \gamma)$ (upper panel) and ${\rm Log}_{10}\Delta a_\mu$ (bottom panel)
in the $(g_{0S \, {\rm or} \,1S},M_{\rm mirror})$ plane for both normal (left panel) and inverted (right panel) neutrino 
mass hierarchies for scenarios 1 (red curves) and 2 (blue curves) with the six cases of couplings aforementioned: 
(1) $g_{0S} \neq 0, \; g_{1S} = 0$ (Fig.~\ref{fig4}),
(2) $g_{1S} = 10^{-2} \times g_{0S}$ (Fig.~\ref{fig5}), 
(3) $g_{1S} = 10^{-1} \times g_{0S}$ (Fig.~\ref{fig6}), 
(4) $g_{1S} = 0.5 \times g_{0S}$ (Fig.~\ref{fig7}), 
(5) $g_{0S} = g_{1S}$ (Fig.~\ref{fig8}),  and
(6) $g_{0S} = 0, \; g_{1S} \neq 0$ (Fig.~\ref{fig9}),
respectively. 

At the upper panel of each of these figures, 
the (light) gray area is excluded by the current limit of 
${\rm Log}_{10}B(\mu \to e \gamma) = -12.24$ from MEG experiment \cite{Adam:2013mnn} 
for scenario (1) 2 respectively.
The projected sensitivity of ${\rm Log}_{10}B(\mu \to e \gamma)=-13.40$ \cite{Renga:2014xra}
is also shown for each scenario in the two plots in the upper panel for comparison.

At the bottom panel of each of these figures, 
the red (blue) area is defined by the 
${\rm Log}_{10}\Delta a_\mu = -8.54$ \cite{muonanomaly} 
from the E821 experiment of the Brookhaven National Lab (BNL) 
for the discrepancy between the SM model prediction 
and the measurement for the muon anomalous magnetic dipole moment
for scenario 1 (2), respectively. 

From all the plots in these figures, we observe the following general features.

\begin{itemize}

\item
In the same mass range of the mirror leptons 
the LFV process $\mu \to e \gamma$ is more sensitive to the couplings
by almost two order of magnitudes as compared with the anomalous magnetic dipole moment of the muon.
This is partly due to the fact that the $B(\mu \to e \gamma)$ is quartic in the couplings, while in $\Delta a_\mu$ 
they are quadratic. 

\item
As one turns on the $A_4$ triplet coupling $g_{1S}$ from 0 to $g_{1S} = g_{0S}$ (Fig.~(\ref{fig4}) to Fig.~(\ref{fig8})),
the contours for ${\rm Log}_{10}B(\mu \to e \gamma)$ (upper panels) are shifting toward to the left, indicating 
the role of the triplet singlets become more relevant and thus the constraints on parameter space become more
stringent from the current MEG limit. However in the last case of Fig.~(\ref{fig9}) 
when the $A_4$ singlet coupling $g_{0S}$ is set to zero
such that only the triplet singlets are contributing in the loop diagram, the contours of ${\rm Log}_{10}B(\mu \to e \gamma)$ 
are slightly shifting back toward to the right. 
Similar behaviors can be found for the contours of ${\rm Log}_{10}\Delta a_\mu$, 
but the effects are tiny and not easily seen on the log scale, except for the last three cases 
of Figs.~(\ref{fig7})-(\ref{fig9}) (lower panels).

\end{itemize}

Regarding the sensitivity on the two scenarios, we can obtain the following statement  
by comparing the red and blue contours corresponding to the scenarios 1 and and 2 
in each of these figures.

\begin{itemize}

\item
The sensitivity of the couplings in the $B(\mu \to e \gamma)$ has been weakened by one to two order of magnitudes
for scenario 2 as compared to scenario 1. This is due to the fact that in scenario 2, the three unknown unitary mixing matrices are now departure from $U^\dagger_{CW}$, which allows the couplings take on larger values since the amplitudes involve products of the couplings and the elements of mixing matrices.
However this sensitivity is not present for the muon anomalous magnetic dipole moment as the distance between
the two red and blue contours for the two scenarios in the lower panels of all these plots 
are well within a small range of the coupling $g_{0S}$ (or $g_{1S}$ in  Fig.~(\ref{fig9})).
For example, at $M_{\rm mirror} = 100$ GeV, the allowed value of $g_{0S}$ varies from $10^{-4.5}$  to 
$10^{-1.8}$ ($10^{-1.9}$ to $10^{-1.4}$) as seen from the upper (lower) panels of Figs.(\ref{fig4})-(\ref{fig8}).

\end{itemize}

Regarding the sensitivity on the neutrino mass hierarchies, one can obtain the following statements
by comparing the left and right panels in each of these figures.

\begin{itemize}

\item
As one slowly turns on the $A_4$ triplet coupling $g_{1S} = 0$ (Fig.~(\ref{fig4}))
to $g_{1S} = 10^{-1} \times g_{0S}$ (Fig.~(\ref{fig6})), the red contours of   
${\rm Log}_{10}B(\mu \to e \gamma)$ of scenario 1 in the left and right panels in all these plots remain the same, 
while the blue contours of scenario 2 in the right panels move toward to the left. 
This indicates that noticeable differences in the contours of ${\rm Log}_{10}B(\mu \to e \gamma)$ 
between the normal and inverted neutrino mass hierarchies can be seen in these cases. 
In general the couplings are about an order of magnitude more sensitive in the inverted mass hierarchy 
than the normal one for scenario 2. However, for $g_{1S} \geq 0.5 \times g_{0S}$, these differences diminish.

\item
There are no discernible differences between the two mass hierarchies 
for the muon anomalous magnetic dipole moment in both scenarios for all 6 cases of couplings.

\end{itemize}

\section{Implications}

The constraints on the Yukawa couplings coming from $\mu \rightarrow e \gamma$ has several implications among which two are particularly relevant.

\begin{itemize}

\item The size allowed for the Yukawa couplings by present limits on $B(\mu \to e \gamma)$ has an important implication on the decay lengths of the mirror leptons. It is beyond the scope of this paper to discuss this in detail here but a few remarks are in order. In the search for mirror leptons, one would like to look for characteristic signatures which can be distinguished from SM background. One of such signatures could be events with displaced vertices, in particular events with decay lengths which are macroscopic ($l > 1\, {\rm mm}$). How this type of events can be correlated to $\mu \rightarrow e \gamma$ is a topic which was already mentioned in \cite{Hung:2007ez}. With the present update which includes a more detailed analysis taking into account mixings in the lepton sector, one can have a better idea of the correlation between the feasibility to observe $\mu \rightarrow e \gamma$ and the detection of mirror leptons. 

A mirror lepton can decay directly into SM leptons with an accompanying Higgs singlet. For example, one can have $l^{M}_{Ri} \rightarrow l_{Lj} + \phi_{kS}$ where $i,j=e,\mu,\tau$ and $k=0,1,2,3$. The decay length will depend on the magnitude of the Yukawa couplings as well as on the various mixing parameters contained in Eq.~(\ref{totalLS}). We just take one example here for the sake of discussion. The interaction Lagrangian for $\mu^{M}_{Ri} \rightarrow l_{Lj} + \phi_{kS}$ can be expressed as $(\bar{e}_L \mathcal{M}_{12} + \bar{\mu}_L \mathcal{M}_{22} + \bar{\tau}_L \mathcal{M}_{32}) \mu_R^M$ where (for scenario 2 with the normal hierarchy)
\bea
\label{couplingyuk}
\mathcal{M}_{12} &=& (5.834 \times 10^{-6} - 0.000025 i) g_{0S} \phi_{0S} + \\ \nonumber
 &&(g_{1S} (0.324 + 0.159 i) + g_{2S} (0.407 - 0.171i)) \phi_{1S} + \\ \nonumber
 &&(g_{1S} (0.154 + 0.200 i) + g_{2S}(0.192 + 0.238i)) \phi_{2S} + \\ \nonumber
 &&(g_{1S} (0.074 - 0.325 i) + g_{2S} (0.201 - 0.102i)) \phi_{3S}  \\ \nonumber  
\mathcal{M}_{22} &=& 0.999933 g_{0S} \phi_{0S} +  \\ \nonumber 
 &&(g_{1S} (-0.262 + 0.332 i) + g_{2S} (-0.262 - 0.332i)) \phi_{1S} +  \\ \nonumber 
 &&(g_{1S} (0.146 - 0.193 i) + g_{2S}(0.146 + 0.193i)) \phi_{2S} +  \\ \nonumber 
 &&(g_{1S} (0.067 - 0.255 i) + g_{2S} (0.067 + 0.255i)) \phi_{3S}  \\ \nonumber 
 \mathcal{M}_{32} &=& (0.00006 + 0.00002 i) g_{0s} \phi_{0S} +  \\ \nonumber 
 &&(g_{1S} (-0.054 - 0.276 i) + g_{2S} (-0.145 + 0.257i)) \phi_{1S} +  \\ \nonumber 
 &&(g_{1S} (-0.163 - 0.043 i) + g_{2S}(0.269 + 0.405i)) \phi_{2S} +  \\ \nonumber 
 &&(g_{1S} (0.166 - 0.503 i) + g_{2S} (-0.157 - 0.077i)) \phi_{3S}  
\eea
Depending on the particular search ($e$, $\mu$ or $\tau$), a displaced vertex might occur. For instance, if one focuses on $\tau$,  and if $g_{iS} \ll g_{0S}$, the constraint on $g_{0S} < 10^{-3}$ (see the above figures) implies that $\mu^{M}_{Ri} \rightarrow \tau_{L} + \phi_{kS}$ would have a macroscopic decay length. There are many such cases but it is beyond the scope of this paper to discuss this issue at length. We merely point out the relationship between the constraints coming from $\mu \rightarrow e \gamma$ and the implication on the search for mirror leptons.

\item The other implication concerns the VEV of the singlet Higgs fields. Since the seesaw mechanism implies the masses of the light neutrinos are given by $\sim m_D^2/M$ and with $M \sim O(\Lambda_{EW})$, it was stated in \cite{Hung:2006ap} that $m_D \sim O(100 \kev)$ and that the singlet VEV $\sim O(100 \kev)$ if $g_{S} \sim O(1)$. However, constraints from $\mu \rightarrow e \gamma$  imply $g_{0S} < 10^{-3}$ which now brings the singlet VEV up to $O(100 \mev)$. In fact it can even be of the order $O(1 \gev)$. From this observation, it is safe to say that there does not appear to be much of a hierarchy problem in the EW-scale $\nu_R$ model.
\end{itemize}


\section{Conclusions}

In this work, we present an update on a previous analysis \cite{Hung:2007ez} for the 
process $\mu \to e \gamma$ performed in the original EW-scale $\nu_R$  model \cite{Hung:2006ap} to an extended
model \cite{Hoang:2014pda}. Mixings effects of neutrinos and charged leptons constructed with 
a $A_4$ symmetry as recently studied in \cite{Hung:2015nva} are also taken into account. 
In this context, the rare process $\mu \to e \gamma$ is link to interesting new physics beyond the SM 
in the lepton sector, like neutrino and charged lepton mass mixings, neutrino mass hierarchies, mirror leptons as well as 
singlet and triplet scalars of $A_4$, etc. The related muon anomalous magnetic dipole moment is also studied in detail for the model.

To summarize, we find that 
\begin{itemize}
\item 
One can deduce more stringent constraints on the parameter space of the EW-scale $\nu_R$
model by using the LFV process $\mu \to e \gamma$  than the muon anomalous magnetic dipole moment.
\item
The branching ratio $B(\mu \to e \gamma)$ shows some sensitivity to the 
neutrino mass hierarchies in scenario 2 but not scenario 1, depending on the 
$A_4$ triplet coupling constants. 
However we are not advocating the use of the process $\mu \to e \gamma$ to settle the issue of neutrino 
mass hierarchies. After all, this is a rare process.
\item
More stringent constraints can be deduced in scenario 1 than scenario 2 using
$B(\mu \to e \gamma)$.
\item
Future data from MEG experiment with the projected sensitivity will impose further constraints on the parameter space of the model.
\item
The muon anomalous magnetic dipole moment is sensitive neither to the neutrino 
mass hierarchies nor the scenarios for all 6 cases of the couplings studied here for the model.
\end{itemize}

Searching for new physics via rare processes is complementary to direct production of new particles at colliders. 
For $\mu \to e \gamma$, the relevant new particles in the model 
are the mirror leptons and scalar singlets running inside the loop diagram. 
As shown in our analysis, the Yukawa couplings of the Higgs singlets to the leptons in the EW-scale $\nu_R$ model
are constrained to be small in order to be consistent with the current experimental limit on $B(\mu \to e \gamma)$. 
Thus searching 
for mirror particles of this model at the LHC would be quite interesting since, due to small couplings, 
they might decay outside the beam pipe and inside the silicon vertex detectors. The $A_4$ singlet and triplet 
scalars are likely to escape detection as missing energies.

As an outlook, one would like to generalize this work to $\mu - e$ conversion.
This work is now in progress and will be reported elsewhere \cite{mueconversion}.

\section*{Appendix}

For the general case of retaining the external fermion masses $m_{i,j}$, the integrals ${\cal I} \left( \frac{m^2_{\phi_{kS}}}{m^2_{{l^M_m}}} \right)$ and
${\mathcal J} \left( \frac{m^2_{\phi_{kS}}}{m^2_{{l^M_m}}} \right)$ in Eqs.~(\ref{CL})-(\ref{CR}) have to be replaced by
${\mathcal I} \left( \frac{m^2_{\phi_{kS}}}{m^2_{{l^M_m}}} , \frac{m^2_i}{m^2_{{l^M_m}}} , \frac{m^2_j}{m^2_{{l^M_m}}} \right)$ and
${\mathcal J} \left( \frac{m^2_{\phi_{kS}}}{m^2_{{l^M_m}}}, \frac{m^2_i}{m^2_{{l^M_m}}} , \frac{m^2_j}{m^2_{{l^M_m}}} \right)$
respectively, where
\begin{eqnarray}
{\cal I}(r,r_i,r_j) & = &  \int_0^1 dx \int_0^{1-x} dy \frac{x ( 1 - x -y) }{x+y +(1-x-y) (r - x r_j - y r_i ) - i 0^+} \; , \nonumber \\
{\cal J}(r,r_i,r_j) & = &  \int_0^1 dx \int_0^{1-x} dy \frac{x + y }{x+y +(1-x-y) (r - x r_j - y r_i ) - i 0^+} \; . \nonumber
\end{eqnarray}


\section*{Acknowledgments}
We would like to thank the hospitality of The International Center of Interdisciplinary Science Education (ICISE)
at Quy Nhon, Vietnam, where this project was completed. TL would like to thank the hospitality of the Institute of Physics, Academia Sinica, Taiwan where part of this project was carried out.
This work was supported in part by the Ministry of Science and Technology (MoST) of Taiwan under
grant numbers 101-2112-M-001-005-MY3 and 104-2112-M-001-001-MY3, by
US DOE grant DE-FG02-97ER41027 and by the Pirrung Foundation.


\end{document}